\documentclass[aps,prd,12pt,preprint,superscriptaddress,preprintnumbers,nofootinbib,floatfix,longbibliography]{revtex4-1}

\usepackage[T1]{fontenc}
\usepackage[utf8]{inputenc}
\usepackage[english]{babel}
\usepackage{amsmath,amsthm,amssymb,amsfonts,mathrsfs,amsbsy,bm}
\usepackage{tensor}
\usepackage{slashed}
 \usepackage{mathrsfs,tikz} 
\usepackage{esint}
\usepackage[a4paper, margin=1.6cm]{geometry}
\usepackage{cancel}
\usepackage{graphicx}
\usepackage{multirow}
\usepackage{array}
\usepackage{booktabs}
\usepackage{makecell}
\usepackage{xcolor}
\colorlet{BLUE}{blue}
\usepackage{tikz}
\usetikzlibrary{quotes,angles,arrows,decorations.markings,decorations.pathmorphing}
\usepackage{hyperref}
\hypersetup{colorlinks=true,breaklinks=true,citecolor=blue,linkcolor=[rgb]{0,0.5,0.9},urlcolor=blue}
\graphicspath{{figures/}}

\newcommand{\be}{\begin{equation}}
\newcommand{\ee}{\end{equation}}
\newcommand{\Be}{\begin{eqnarray}}
\newcommand{\Ee}{\end{eqnarray}}

\newcommand{\mincir}{\raise-3.truept\hbox{\rlap{\hbox{$\sim$}}\raise4.truept\hbox{$<$}\ }}
\newcommand{\magcir}{\raise-3.truept\hbox{\rlap{\hbox{$\sim$}}\raise4.truept\hbox{$>$}\ }}

\providecommand{\U}[1]{}

\newcommand{\ie}{\begin{equation}}
\newcommand{\fe}{\end{equation}}
\newcommand{\se}{\begin{eqnarray}}
\newcommand{\ff}{\end{eqnarray}}

\begin{document}
\emergencystretch=4em
\usetikzlibrary{arrows.meta,patterns}


\title{An exact rotating black hole solution in Palatini Kalb--Ramond gravity}

\author{A. A. Ara\'{u}jo Filho}
\email{dilto@fisica.ufc.br}
\affiliation{Departamento de F\'isica, Universidade Federal da Para\'iba, Caixa Postal 5008, 58051--970, Jo\~ao Pessoa, Para\'iba, Brazil.}
\affiliation{Departamento de F\'isica, Universidade Federal de Campina Grande, Caixa Postal 10071, 58429--900 Campina Grande, Para\'iba, Brazil.}
\affiliation{Center for Theoretical Physics, Khazar University, 41 Mehseti Street, Baku, AZ-1096, Azerbaijan.}


\date{\today}


\begin{abstract}

We construct an exact stationary axisymmetric vacuum configuration in metric affine Kalb--Ramond gravity with zero cosmological constant. Independent variation of the metric, connection, and antisymmetric field yields the complete field equations and the algebraic relation between the physical and auxiliary geometries. A Kerr auxiliary metric and a closed corotating vacuum two form of fixed positive norm determine the physical metric. The tensor kernel contains the horizon generator, and the metrics, independent connection, and vacuum field share a smooth extension across the nonextremal physical Killing horizon and the rotation axes at finite radius for unequal deformation factors. We obtain an explicit sufficient domain for a causal exterior and an asymptotically locally flat end at fixed nonpolar angles. The physical Hamiltonian yields the timelike and null equations, equatorial effective potentials, circular orbit properties, radial stability, and perihelion advance. Numerical photon evolution determines the critical curve on an orthonormal observer screen. Its size, shape, and centroid depend on $\epsilon=\xi_2b^2/[2(1+\xi_1b^2)]$ at fixed normalized mass scale and spin, with Kerr recovered at $\epsilon=0$. Geometric observations probe this combination and retain an exact degeneracy between the curvature couplings. A static equatorial comparison with Mercury gives a conditional sensitivity of $2.30\times10^{-11}$ for $|2\epsilon|$ under the stated angular convention and mass setting.
\end{abstract}

\maketitle

\tableofcontents


\section{Introduction}
\label{sec:introduction}

The metric affine formulation of gravity assigns independent roles to the metric and the connection. The metric determines lengths and causal structure, while the connection governs parallel transport and curvature. Their independent variation accommodates torsion and nonmetricity and allows matter with intrinsic spin, dilation, or shear to act as a source of affine geometry~\cite{Hehl1995MAG,SotiriouLiberati2007MAG}. This framework contains theories with propagating connection degrees of freedom and theories in which the connection is determined by the remaining fields. The distinction is central to the construction and interpretation of exact solutions~\cite{Olmo2011Palatini,BeltranJimenez2018BI}.

Applications of this formalism extend from cosmology to compact objects. Palatini theories have been used to study nonsingular cosmological evolution~\cite{BarraganOlmo2010Bounce}, stellar structure~\cite{Wojnar2019Polytropic}, and charged black hole geometries with modified central regions~\cite{OlmoRubiera2012BH}. Cosmological models with matter carrying hypermomentum further display the dynamical role of the connection~\cite{Iosifidis2020Hyperfluid}. The propagation of tensor perturbations on nonsingular cosmological backgrounds has also been analyzed, with stability depending on the evolution of the auxiliary and physical geometries~\cite{BeltranJimenez2017GW}.

An auxiliary geometry gives rise to a useful route to solving a class of these theories. For suitable actions built from the symmetric Ricci tensor, the connection equation can be expressed as compatibility with an auxiliary metric. The resulting equations take an Einstein form, while an algebraic transformation relates the auxiliary and physical metrics~\cite{Afonso2018Mapping,Afonso2018NEDMapping}. This construction permits known geometries to be used as starting points and makes explicit which metric enters the motion of minimally coupled matter. Nonminimal couplings to a vacuum tensor introduce directional dependence into the transformation, connecting the affine description with spontaneous Lorentz symmetry breaking.

A dynamical tensor can acquire a nonzero vacuum expectation value through a potential and thereby select preferred directions in spacetime~\cite{KosteleckySamuel1989,Altschul2010}. The bumblebee model realizes this mechanism through a vector, whereas the Kalb--Ramond field is an antisymmetric tensor of rank two. Its origin in the interaction of extended objects~\cite{KalbRamond1974} and its appearance in the bosonic sector of string theory provide a separate motivation for studying its gravitational couplings. The rotating charged solution contains a nontrivial antisymmetric field together with a dilaton and an electromagnetic field~\cite{Sen1992}. The Lorentz violating Kalb--Ramond models considered here employ a potential fixing the tensor norm and nonminimal curvature couplings, giving their vacuum sector a distinct gravitational structure~\cite{Altschul2010}.

Black hole solutions in the metric formulation of this sector have developed along several directions. It was obtained a static geometry with a power law correction associated with the Kalb--Ramond vacuum~\cite{Lessa2020KR}. Subsequent work derived Schwarzschild type solutions with and without a cosmological constant~\cite{Yang2023KR}, electrically charged configurations~\cite{Duan2024ChargedKR}, and more general neutral solutions~\cite{Liu2024NeutralKR}. More recent studies have examined solutions in different dimensions, topological horizons, dyonic charges, and the simultaneous presence of two nonminimal curvature couplings~\cite{Liu2025ExactKR,Yu2026RevisitingKR,Lin2026DyonicKR,Yang2026TwoCouplingKR}. A noncommutative extension has also been constructed and examined through its thermal and optical properties~\cite{Araujo2025NoncommutativeKR,Araujo2026NoncommutativeOpticsKR}.

Rotation has also been investigated. Proposed rotating geometries with Kalb--Ramond corrections have been used to study horizons, photon trajectories, and shadows~\cite{Kumar2020RotatingKR}. A solution derived from the field equations to first order in rotation shows a complementary description with a specified vacuum tensor configuration~\cite{Liu2025SlowRotationKR}. Rotating extensions of charged geometries have also been analyzed through shadow observables and Event Horizon Telescope data~\cite{Nengroo2026RotatingKR}. The construction of an exact stationary configuration at finite rotation requires the metric and the antisymmetric field to satisfy their equations together; this requirement fundamentally motivates the auxiliary metric approach developed in the present work.

The phenomenology of Kalb--Ramond backgrounds includes orbital dynamics, gravitational lensing, perturbations, and black hole thermodynamics. It have been investigated charged particle motion and light deflection~\cite{AlBadawi2024KRGeodesics}, thermodynamic quantities and photon propagation in vacuum and plasma~\cite{Sucu2025KRPhoton,Mangut2025KRLensing}, and anti de Sitter geometries with a global monopole~\cite{Sucu2025AdSKR}. Their studies also include black holes coupled to ModMax electrodynamics~\cite{Sucu2026KRModMax}, quintessence backgrounds~\cite{AlBadawi2025QuintessenceKR}, and dyonic configurations with a cloud of strings and applications to quasiperiodic oscillations~\cite{Ahmed2026DyonicKR}.

Wave propagation and image formation provide further applications. Quasinormal spectra and greybody factors have been studied for static Kalb--Ramond backgrounds~\cite{Guo2024QNMKR}, while scalar clouds and quasinormal ringing have been examined on proposed rotating geometries~\cite{Tokgoz2026KRClouds}. Recent calculations consider polarized images produced by thick accretion disks~\cite{Yang2026KRPolarization} and particle dynamics, shadows, and Hawking emission for nonlinear electromagnetic sources~\cite{Ahmed2026NEDKR}.

On the other hand, the metric affine bumblebee theory supplies a direct geometric precedent for this investigation. Its classical formulation expresses the independent connection through a metric related algebraically to the preferred vector, and its quantum treatment exhibits induced interactions with matter~\cite{Delhom2021Classical,Delhom2022Radiative}. Within the traceless gravitational sector, it was obtained a static vacuum solution~\cite{Araujo2023} and subsequently an exact stationary and axisymmetric configuration~\cite{Araujo2024}. The auxiliary Schwarzschild and Kerr geometries made it possible to determine the physical metrics after specifying admissible vector profiles. The resulting solutions were examined through geodesics, classical gravitational tests, and thermodynamics. A more general static configuration with temporal and radial vacuum components has recently extended this construction~\cite{Araujo2026GeneralAffine}.

Several applications have followed these metric affine solutions. Accretion disks, light deflection, greybody bounds, and neutrino propagation were studied in Ref.~\cite{Lambiase2023Affine}. Scalar perturbations, quasinormal frequencies, scattering, absorption, and Hawking radiation have been investigated in Refs.~\cite{Araujo2024Traces,JhaRahaman2024Affine,Heidari2024AffineScattering}. Further studies concern gravitational lensing~\cite{Gao2024Affine,Araujo2025AffineLensing}, shadows and quasiperiodic oscillations~\cite{JhaRahaman2025Affine}, and the images of rotating configurations~\cite{Nascimento2026AffineShadows}. Gravitational wave propagation and emission have also been analyzed in the corresponding linearized theory~\cite{Araujo2026AffineWaves}.

A recent development in this direction is the static and spherically symmetric family studied Palatini Kalb--Ramond gravity~\cite{She2026}. Their analysis considered a pseudo electric vacuum, both nonminimal curvature couplings, and different values of the cosmological constant, with applications to conserved quantities, thermodynamics, and Solar System tests. For a simple two form, the quadratic curvature coupling acts on a two dimensional plane, changing the algebraic reconstruction from that of a vector vacuum. In this manner, extending the construction to stationary and axisymmetric geometries requires the metric transformation before spherical symmetry is imposed and a tensor profile adapted to the rotating auxiliary geometry. Its fixed norm and closure conditions determine which profiles are admissible therefore.

In this work, we derive that transformation and construct an exact stationary axisymmetric vacuum configuration with vanishing cosmological constant. We choose Kerr as the auxiliary metric and determine a simple, closed corotating Kalb Ramond two form of fixed positive norm. Its temporal components align the tensor kernel with the horizon generator, preserving a regular physical Killing horizon for unequal deformation factors. The physical metric, auxiliary metric, independent connection, and vacuum field share a smooth extension across the nonextremal horizon and the rotation axes at finite radius. We verify the connection, metric, and tensor equations and obtain an explicit sufficient parameter domain for a causal exterior. We also characterize the asymptotically locally flat end at fixed nonpolar angles, retaining the original $2\pi$ azimuthal period required by axis regularity.

We then derive the timelike and null equations from the physical Hamiltonian, retaining the mixed metric components and the coupled radial and polar motion. On the equatorial plane, we obtain the effective potentials, circular orbit energies and angular momenta, radial stability conditions, and perihelion advance. Numerical photon evolution determines the critical curve on an observer's orthonormal screen. At fixed normalized mass scale, spin, and inclination, its size, shape, and displacement vary with $\epsilon=\xi_2b^2/[2(1+\xi_1b^2)]$, while $\epsilon=0$ reproduces Kerr. We relate the calculated shadow diameters to the published measurements of M87* and Sgr A*~\cite{EHTM872019VI,EHTSgr2022VI}, derive conditional perihelion sensitivities from planetary ephemerides, and formulate the stellar precession comparison using the physical trajectories and received photons. These geometric predictions retain the exact coupling degeneracy $\xi_2b^2=2\epsilon(1+\xi_1b^2)$ at fixed normalized mass scale and spin.


\section{Metric affine Kalb--Ramond gravity and the auxiliary geometry}
\label{sec:framework}

The independent connection offers a useful route to exact solutions in theories with spontaneous Lorentz symmetry breaking. In this description, the curvature is determined by an auxiliary metric, whereas the physical metric follows from an algebraic relation involving the vacuum field. This construction was employed in the bumblebee setting in Ref.~\cite{Araujo2024}. Here, we develop its counterpart for an antisymmetric Kalb--Ramond field, using the gravitational couplings considered in Ref.~\cite{She2026}. Our purpose is to obtain the relation between the two metrics without imposing spherical symmetry, so that the same formalism can be applied to a stationary and axisymmetric geometry.


\subsection{The model and its independent variables}

We consider a four dimensional spacetime with signature $(-,+,+,+)$ and action
\begin{equation}
\begin{split} S = {}&\frac{1}{2\kappa^2}\int \mathrm{d}^4x\,\sqrt{-g}\,\bigg[(1+\xi_1N)R(\Gamma)+\xi_2B^{\mu\alpha}B^\nu{}_{\alpha}R_{(\mu\nu)}(\Gamma)-\frac{1}{6}H_{\mu\nu\rho}H^{\mu\nu\rho}-V(X)\bigg]+S_{\mathrm{mat}}[g,\psi], \end{split}
\label{eq:action}
\end{equation}
where $\kappa^2 = 8\pi G$, and $\xi_1$ and $\xi_2$ control the nonminimal interactions. We take the cosmological constant to vanish. The metric $g_{\mu\nu}$, the symmetric connection $\Gamma^\lambda{}_{\mu\nu}$, and the antisymmetric field $B_{\mu\nu}$ are varied independently. It is important to mention that ordinary matter couples to $g_{\mu\nu}$ and does not depend on $\Gamma^\lambda{}_{\mu\nu}$. In addition, throughout this section, indices are raised and lowered with $g_{\mu\nu}$ unless another metric is displayed explicitly. We define
\begin{equation}
N = B_{\mu\nu}B^{\mu\nu},\qquad X = N-\sigma b^2,\qquad \sigma = \pm1,\qquad R(\Gamma) = g^{\mu\nu}R_{(\mu\nu)}(\Gamma),
\label{eq:definitions}
\end{equation}
with $b^2 > 0$. The sign $\sigma$ distinguishes positive and negative vacuum norms. The purely spatial configuration considered in the application below has $\sigma = + 1$. Antisymmetrization carries unit weight; in particular,
\begin{equation}
H_{\mu\nu\rho} = \partial_{[\mu}B_{\nu\rho]}=\frac{1}{3}\left(\partial_\mu B_{\nu\rho} + \partial_\nu B_{\rho\mu}+\partial_\rho B_{\mu\nu}\right).
\label{eq:fieldstrength}
\end{equation}
This convention fixes the numerical factors in the field equation below. In differential form notation, with $\bm B = \tfrac12 B_{\mu\nu}\mathrm{d}x^\mu\wedge\mathrm{d}x^\nu$ and $\bm H = \tfrac1{3!}H_{\mu\nu\rho}\mathrm{d}x^\mu\wedge\mathrm{d}x^\nu\wedge\mathrm{d}x^\rho$, it gives $\bm H = \tfrac13\,\mathrm{d}\bm B$. The field strength contains no independent connection, as it should be expected.

The vacuum expectation value is denoted by $\langle B_{\mu\nu}\rangle = b_{\mu\nu}$ and obeys
\begin{equation}
g^{\mu\alpha}g^{\nu\beta}b_{\mu\nu}b_{\alpha\beta} = \sigma b^2.
\label{eq:physicalnorm}
\end{equation}
For the vacuum construction below, we assume that the smooth potential has a minimum at $X = 0$, with $V(0) = V'(0) = 0$. The norm condition is imposed after varying the action: the metric dependence of $N$ must be retained in the variation.


\subsection{The connection equation and the metric transformation}

It is convenient to collect the curvature couplings into the symmetric tensor
\begin{equation}
K^{\mu\nu} = (1+\xi_1N)g^{\mu\nu} + \xi_2B^{\mu\alpha}B^\nu{}_{\alpha},\qquad P^{\mu\nu} = \sqrt{-g}\,K^{\mu\nu}.
\label{eq:K}
\end{equation}
Here, $P^{\mu\nu}$ is a symmetric tensor density of weight $+1$, and its covariant derivative includes the corresponding density term. The curvature part of the action is $P^{\mu\nu}R_{\mu\nu}(\Gamma)/(2\kappa^2)$. Varying the symmetric connection and integrating by parts gives
\begin{equation}
\nabla^{(\Gamma)}_\lambda P^{\mu\nu} - \delta_\lambda^{(\mu}\nabla^{(\Gamma)}_\rho P^{\nu)\rho} = 0.
\label{eq:connectionvariation}
\end{equation}
Contracting $\lambda$ with $\mu$ yields $-\tfrac32\nabla^{(\Gamma)}_\rho P^{\nu\rho}=0$ in four dimensions. In this sense, the connection equation reduces to
\begin{equation}
\nabla^{(\Gamma)}_\lambda P^{\mu\nu} = 0.
\label{eq:densitycompatibility}
\end{equation}
For a nondegenerate $K^{\mu\nu}$ of Lorentzian signature, we introduce $h_{\mu\nu}$ through
\begin{equation}
\sqrt{-h}\,h^{\mu\nu} = \sqrt{-g}\,K^{\mu\nu}, \qquad h = \det(h_{\mu\nu}),\quad g = \det(g_{\mu\nu}).
\label{eq:auxdefinition}
\end{equation}
Eq.~\eqref{eq:densitycompatibility} now determines the independent connection as the Levi Civita connection of $h_{\mu\nu}$:
\begin{equation}
\Gamma^\lambda{}_{\mu\nu} = \frac12h^{\lambda\rho}\left(\partial_\mu h_{\rho\nu} + \partial_\nu h_{\rho\mu} - \partial_\rho h_{\mu\nu}\right).
\label{eq:LC}
\end{equation}

To make the transformation explicit, define the mixed tensor
\begin{equation}
\mathcal M^\mu{}_{\nu} = K^{\mu\alpha}g_{\alpha\nu} = (1+\xi_1N)\delta^\mu{}_{\nu} + \xi_2B^{\mu\alpha}B_{\nu\alpha},\qquad D = \det\mathcal M.
\label{eq:M}
\end{equation}
Taking determinants of Eq.~\eqref{eq:auxdefinition} gives
\begin{equation}
\det\!\left(\sqrt{-h}\,h^{\mu\nu}\right) = h,\qquad \det\!\left(\sqrt{-g}\,K^{\mu\nu}\right) = g D.
\end{equation}
Thereby, $h = gD$ and $\sqrt{-h}=\sqrt{-g}\sqrt D$. Substituting these expressions back into Eq.~\eqref{eq:auxdefinition}, and then taking the matrix inverse, we obtain
\begin{align}
h^{\mu\nu}& = \frac{1}{\sqrt D}\,\mathcal M^\mu{}_{\alpha}g^{\alpha\nu},\label{eq:generalinverse}\\ h_{\mu\nu}& = \sqrt D\,g_{\mu\alpha}(\mathcal M^{-1})^\alpha{}_{\nu}.\label{eq:generalcovariant}
\end{align}
These relations require $D > 0$ and preservation of the Lorentzian signature. A remarke here is imporatnt: they hold for an arbitrary antisymmetric field and precede any choice of coordinates or spacetime symmetry. Since $\mathcal M^\mu{}_{\nu}$ itself contains $g_{\mu\nu}$, reconstruction from a prescribed auxiliary metric needss an algebraic inversion. We give its explicit form for a simple vacuum two form in what follows.

Furthermore, the role of the auxiliary geometry can also be seen directly in the action:
\begin{equation}
\sqrt{-g}\,K^{\mu\nu}R_{\mu\nu}(\Gamma) = \sqrt{-h}\,h^{\mu\nu}R_{\mu\nu}(h).
\label{eq:curvatureidentity}
\end{equation}
In this manner, its curvature term takes the Einstein Hilbert form. The remaining terms still depend on the physical metric through the field contractions and the volume element. In particular, the two scalar curvatures must be distinguished:
\begin{equation}
R(\Gamma) = g^{\mu\nu}R_{\mu\nu}(h), \qquad R[h] = h^{\mu\nu} R_{\mu\nu}(h).
\label{eq:twoscalars}
\end{equation}


\subsection{The metric and Kalb--Ramond equations}

We next vary $g^{\mu\nu}$ while holding $B_{\mu\nu}$ and $\Gamma^\lambda{}_{\mu\nu}$ fixed. To display the result without absorbing curvature terms into an effective matter tensor, let us introduce
\begin{equation}
Z_{\mu\nu} = B_{\mu\alpha}B_\nu{}^\alpha,\qquad Q^{\mu\nu} = B^{\mu\alpha}B^\nu{}_{\alpha}, \qquad \mathcal R_{\mu\nu} = R_{(\mu\nu)}(\Gamma).
\end{equation}
The variation of the norm is $\delta N = 2Z_{\mu\nu}\delta g^{\mu\nu}$. The metric dependence of $Q^{\alpha\beta}$ gives us
\begin{equation}
\begin{split} \mathcal C_{\mu\nu}\equiv{}& \frac{\partial Q^{\alpha\beta}}{\partial g^{\mu\nu}}\mathcal R_{\alpha\beta}={} B^{\alpha\beta}B_{\mu\beta}\mathcal R_{\nu\alpha} + B^{\alpha\beta}B_{\nu\beta}\mathcal R_{\mu\alpha}+B^\alpha{}_{\mu}B^\beta{}_{\nu}\mathcal R_{\alpha\beta}. \end{split}
\label{eq:C}
\end{equation}
Accordingly, the metric equation reads
\begin{equation}
\begin{split} &(1 + \xi_1N)\mathcal R_{\mu\nu} -\frac12 g_{\mu\nu}\left[(1+\xi_1N)R(\Gamma) + \xi_2Q^{\alpha\beta}\mathcal R_{\alpha\beta}\right]\\ &\hspace{8mm}  +  2\xi_1Z_{\mu\nu}R(\Gamma) + \xi_2\mathcal C_{\mu\nu} = \kappa^2T^{\mathrm{mat}}_{\mu\nu}+\Theta_{\mu\nu}, \end{split}
\label{eq:metricequation}
\end{equation}
where
\begin{align}
T^{\mathrm{mat}}_{\mu\nu}& = -\frac{2}{\sqrt{-g}}\frac{\delta S_{\mathrm{mat}}}{\delta g^{\mu\nu}},\\ \Theta_{\mu\nu}& = \frac12H_{\mu\alpha\beta}H_\nu{}^{\alpha\beta}-\frac1{12}g_{\mu\nu}H_{\alpha\beta\gamma}H^{\alpha\beta\gamma} + 2V'Z_{\mu\nu}-\frac12g_{\mu\nu}V.\label{eq:Theta}
\end{align}
The normalization of $\Theta_{\mu\nu}$ follows from placing the Kalb--Ramond terms inside the prefactor in Eq.~\eqref{eq:action}. No derivatives of the curvature couplings arise from the metric variation, since the connection is independent.

Variation with respect to $B_{\mu\nu}$ gives the remaining field equation,
\begin{equation}
\nabla^{(g)}_\alpha H^{\alpha\mu\nu} = -6\xi_1B^{\mu\nu}R(\Gamma) - 6\xi_2B^{\gamma[\nu}\mathcal R^{\mu]}{}_{\gamma} + 6V'B^{\mu\nu}.
\label{eq:KRequation}
\end{equation}
Here, notice that $V'=\frac{\mathrm{d}V}{\mathrm{d}X}$, and $\nabla^{(g)}$ is constructed from the Levi Civita connection of $g_{\mu\nu}$. Eqs.~\eqref{eq:LC}, \eqref{eq:metricequation}, and \eqref{eq:KRequation} determine the geometry and the tensor field.


\subsection{A vacuum two form}

The metric transformation admits a particularly transparent form for a simple vacuum two form, meaning that it is the exterior product of two one forms. Locally, it can be written as
\begin{equation}
b_{\mu\nu} = \frac{b}{\sqrt2}\left(n_\mu e_\nu-n_\nu e_\mu\right),\qquad n_\mu n^\mu = \sigma,\quad e_\mu e^\mu = 1,\quad n_\mu e^\mu = 0,
\label{eq:electricVEV}
\end{equation}
where $b = \sqrt{b^2}$. This choice obeys Eq.~\eqref{eq:physicalnorm}. For $\sigma = +1$, both one forms are spacelike; for $\sigma = - 1$, $n_\mu$ is timelike. It is important to highlith that the choice specifies the algebraic type of the vacuum field only.

The field singles out a two dimensional plane with projector
\begin{equation}
\Pi^\mu{}_{\nu} = \sigma n^\mu n_\nu+e^\mu e_\nu,\qquad \Pi^\mu{}_{\alpha}\Pi^\alpha{}_{\nu} = \Pi^\mu{}_{\nu},\qquad \Pi^\mu{}_{\mu} = 2.
\label{eq:projector}
\end{equation}
Using Eq.~\eqref{eq:electricVEV}, we obtain
\begin{equation}
b^{\mu\alpha}b_{\nu\alpha} = \frac{\sigma b^2}{2}\Pi^\mu{}_{\nu}.
\label{eq:Bsquared}
\end{equation}
Let us define
\begin{equation}
\ell_1 = \xi_1b^2,\qquad \ell_2 = \xi_2 b^2,\qquad A = 1+\sigma\ell_1,\qquad C=A+\frac{\sigma\ell_2}{2}.
\label{eq:AC}
\end{equation}
The deformation tensor then separates into the two complementary planes:
\begin{equation}
\mathcal M^\mu{}_{\nu} = C\,\Pi^\mu{}_{\nu}+A\left(\delta^\mu{}_{\nu}-\Pi^\mu{}_{\nu}\right).
\label{eq:projectorM}
\end{equation}
Each projector has rank two, so the determinant and inverse follow immediately:
\begin{equation}
D  = A^2C^2,\qquad (\mathcal M^{-1})^\mu{}_{\nu} = \frac1C\Pi^\mu{}_{\nu} + \frac1A\left(\delta^\mu{}_{\nu} -\Pi^\mu{}_{\nu}\right).
\label{eq:electricdet}
\end{equation}
We take $A > 0$ and $C > 0$, which preserve the signature and allow a continuous limit to general relativity. With $\Pi_{\mu\nu} = g_{\mu\alpha}\Pi^\alpha{}_{\nu}$, Eq.~\eqref{eq:generalcovariant} becomes
\begin{equation}
h_{\mu\nu} = A\Pi_{\mu\nu}+C(g_{\mu\nu} - \Pi_{\mu\nu}) = Cg_{\mu\nu}-\xi_2b_\mu{}^\alpha b_{\nu\alpha}.
\label{eq:electricmap}
\end{equation}
In other words, the auxiliary metric multiplies the field plane by $A$ and its orthogonal complement by $C$. This explains why the relation can be used without assuming a diagonal physical metric.

For constructing a solution, it is useful to reverse Eq.~\eqref{eq:electricmap}. Since $b_{\mu\nu}$ has support only on the field plane,
\begin{equation}
b_{\mu\alpha}h^{\alpha\beta}b_{\nu\beta} = \frac1A b_{\mu\alpha} g^{\alpha\beta}b_{\nu\beta}.
\end{equation}
In this sense, the physical metric can be reconstructed from the auxiliary metric and the covariant vacuum field as
\begin{equation}
g_{\mu\nu} = \frac1C\left(h_{\mu\nu} + \xi_2A\,b_{\mu\alpha}h^{\alpha\beta}b_{\nu\beta}\right).
\label{eq:reconstruction}
\end{equation}
The corresponding norm expressed entirely with $h_{\mu\nu}$ is
\begin{equation}
h^{\mu\alpha}h^{\nu\beta}b_{\mu\nu}b_{\alpha\beta} = \frac{\sigma b^2}{A^2}.
\label{eq:auxnorm}
\end{equation}
In this manner, one of the most remarkable features in developing this approach is that we may start with $h_{\mu\nu}$, choose a simple field satisfying Eq.~\eqref{eq:auxnorm}, and obtain $g_{\mu\nu}$ from Eq.~\eqref{eq:reconstruction}. The physical norm then has the required value. The separate condition $H_{\mu\nu\rho} = 0$ requires the vacuum two form $\bm b = \tfrac12 b_{\mu\nu}\mathrm{d}x^\mu\wedge\mathrm{d}x^\nu$ to be closed, $\mathrm{d}\bm b=0$; notice that it does not follow from its norm or algebraic type.


\subsection{The vacuum equations and the route to (other) exact solutions}

We now set $T^{\mathrm{mat}}_{\mu\nu} = 0$ and $B_{\mu\nu} = b_{\mu\nu}$. For a closed vacuum field, $H_{\mu\nu\rho} = 0$. With $V = V' = 0$, a sufficient condition for Eqs.~\eqref{eq:metricequation} and \eqref{eq:KRequation} is
\begin{equation}
R_{\mu\nu}(h) = 0.
\label{eq:RicciFlat}
\end{equation}
This provides a direct construction of exact backgrounds. we first choose an auxiliary vacuum geometry, such as Kerr, and then determines a closed vacuum field with the prescribed norm. The physical metric follows algebraically from Eq.~\eqref{eq:generalcovariant}, or from Eq.~\eqref{eq:reconstruction} for a simple vacuum two form. The nonminimal curvature terms vanish because the independent connection is governed by the Ricci flat auxiliary metric. In addition, the physical metric need not itself be Ricci flat with respect to its own Levi Civita connection.


\section{A stationary axisymmetric vacuum with a regular physical horizon}
\label{sec:generalhorizon}

We construct a stationary and axisymmetric vacuum with a simple, closed Kalb--Ramond two form of positive norm. The auxiliary metric is Kerr, as we have mentioned the earlier discussions, and the physical metric follows from the algebraic reconstruction. Temporal components of the vacuum tensor align its kernel with the Kerr horizon generator. This alignment preserves a smooth physical Killing horizon for unequal deformation factors. We retain
\begin{equation}
A=1+\xi_1b^2>0,\qquad
C=A+\frac{\xi_2b^2}{2}>0,
\label{eq:generalhorizonAC}
\end{equation}
with $b^2>0$, $H_{\mu\nu\rho}=0$, and $V=V'=0$. In addition, as we have argued in the previous section, ordinary matter and the cosmological constant vanish. It is important to menttion that positive deformation factors $(A,C)$ preserve the Lorentzian signature of the reconstructed metric. Furthermore, we establish below an explicit sufficient parameter domain for a causal exterior. Here, a very important remark should be highlithed: the existence of a regular physical horizon also depends on the vacuum configuration $b_{\mu\nu}$ and on a common smooth extension of the metric and independent fields. The corotating configuration adopted here satisfies this local horizon requirement for unequal deformation factors. In Sec.~\ref{subsec:alternativeKRconfigurations}, we examine other choices of $b_{\mu\nu}$ and the geometries they generate, assessing their black hole interpretation through the same regularity, causal, and asymptotic criteria.


\subsection{The auxiliary Kerr geometry and a corotating vacuum two form}

In Boyer--Lindquist coordinates $(t,r,\theta,\phi)$, define
\begin{equation}
\Sigma=r^2+a^2\cos^2\theta,\qquad
\Delta=r^2-2mr+a^2,\qquad
r_\pm=m\pm\sqrt{m^2-a^2},\qquad 0<a<m.
\label{eq:generalhorizonKerrfunctions}
\end{equation}
The auxiliary metric is
\begin{equation}
\begin{split}
\mathrm ds_h^2={}&-\frac{\Delta}{\Sigma}
(\mathrm dt-a\sin^2\theta\,\mathrm d\phi)^2
+\frac{\sin^2\theta}{\Sigma}
[(r^2+a^2)\mathrm d\phi-a\,\mathrm dt]^2 +\Sigma\left(\frac{\mathrm dr^2}{\Delta}+\mathrm d\theta^2\right).
\end{split}
\label{eq:generalhorizonKerrmetric}
\end{equation}
Let
\begin{equation}
q=r^2+a^2,\qquad q_+=r_+^2+a^2,\qquad
\Omega_H=\frac{a}{q_+},\qquad
\zeta=\mathrm d\phi-\Omega_H\,\mathrm dt.
\label{eq:generalhorizoncorotation}
\end{equation}
We use $\zeta$ as an angular one form; it is unrelated to the particle normalization parameter used in the geodesic section. Now, let us introduce
\begin{equation}
k=\frac{\varepsilon_b b}{\sqrt2 A},\qquad
\varepsilon_b=\pm1,\qquad
z=\sin^2\theta,\qquad c=\cos\theta,\qquad
x=\frac{q-q_+}{q_+},\qquad d=1+xz,
\label{eq:generalhorizonprofiledefinitions}
\end{equation}
and the scalar flux function
\begin{equation}
F=-\frac{kq_+\cos\theta}{\sqrt d},\qquad
F_r=\frac{kr\sin^2\theta\cos\theta}{d^{3/2}},\qquad
F_\theta=\frac{kq\sin\theta}{d^{3/2}}.
\label{eq:generalhorizonflux}
\end{equation}
The two form is
\begin{equation}
\boldsymbol b=\psi\,\mathrm dr\wedge\mathrm d\theta
+\mathrm dF\wedge\zeta.
\label{eq:generalhorizontwoform}
\end{equation}
With $\boldsymbol b=\tfrac12 b_{\mu\nu}\,\mathrm dx^\mu\wedge\mathrm dx^\nu$, its components are displayed by the antisymmetric matrix
\begin{equation*}
\bigl[b_{\mu\nu}\bigr]_{(t,r,\theta,\phi)}=
\begin{pmatrix}
0 & \Omega_H F_r & \Omega_H F_\theta & 0\\
-\Omega_H F_r & 0 & \psi & F_r\\
-\Omega_H F_\theta & -\psi & 0 & F_\theta\\
0 & -F_r & -F_\theta & 0
\end{pmatrix}.
\end{equation*}
The subscript specifies the coordinate order of both rows and columns.
It is simple because it involves only the three one forms $\mathrm dr$, $\mathrm d\theta$, and $\zeta$. It is closed because $\mathrm d\zeta=0$, $\mathrm d^2F=0$, and $\psi$ depends only on $r$ and $\theta$.

The remaining coefficient is fixed by the auxiliary norm. In this sense, let us define
\begin{equation}
D=\frac{(1-a\Omega_H z)^2}{z}
-\frac{(a-\Omega_H q)^2}{\Delta},\qquad
U=\Delta F_r^2+F_\theta^2.
\label{eq:generalhorizonDU}
\end{equation}
Here, $h^{-1}(\zeta,\zeta)=D/\Sigma$. The fixed norm condition is
\begin{equation}
\frac12 h^{\mu\alpha}h^{\nu\beta}b_{\mu\nu}b_{\alpha\beta}
=\frac{\Delta\psi^2+DU}{\Sigma^2}=k^2,
\qquad
\psi^2=\frac{k^2\Sigma^2-DU}{\Delta}.
\label{eq:generalhorizonnorm}
\end{equation}
The apparent singularities in this expression are removable at $r_+$ and at the axes. An expression suitable for direct evaluation there is obtained by setting
\begin{equation}
\lambda_r=\frac{r+r_+}{r-r_-},\qquad
\beta=\frac{a^2}{q_+},\qquad \alpha=1-\beta z,
\label{eq:generalhorizonregularvariables}
\end{equation}
where $0<\beta<1/2$, and defining
\begin{equation}
\begin{split}
\mathcal H={}&2\alpha\beta+3\alpha^2
+x[2\beta+6\alpha+z(3\alpha^2-\beta^2)]\\
&+x^2(3+6\alpha z+\alpha^2z^2)
+x^3z(3+2\alpha z)+x^4z^2,\\
\mathcal B={}&q_+\lambda_r\mathcal H-\alpha^2r^2c^2
+\frac{a^2\lambda_r^2}{q_+^2}
(q^2+\Delta r^2zc^2).
\end{split}
\label{eq:generalhorizonradicand}
\end{equation}
Then one choice of orientation is
\begin{equation}
\psi=\frac{k\sin\theta\sqrt{\mathcal B}}{d^{3/2}}.
\label{eq:generalhorizonpsi}
\end{equation}
Reversing $k$ reverses the complete two form. The sign of $\psi$ can also be reversed independently in the norm and closure equations.

The radicand is positive throughout the exterior. To see this, it is convenient to introduce
\begin{equation}
W=q^2-a^2\Delta z,\qquad
L_F=q^2+\Delta r^2zc^2,
\label{eq:generalhorizonWL}
\end{equation}
so that $U=k^2zL_F/d^3$. Completing the square in $D$ gives
$D\leq\Sigma^2/(zW)$. Furthermore,
$W\geq q_+\Sigma$ and
\begin{equation}
Wd^3-L_F\geq2\Delta\Sigma z>0
\qquad (r>r_+,\ 0<\theta<\pi).
\label{eq:generalhorizonpositivity}
\end{equation}
In other words,
$\mathcal B\geq2\Sigma^3/W>0$, with the inequality extending to the removable limits. For example, at the horizon,
\begin{equation}
\mathcal B_+=q_+\lambda_+
(2\alpha\beta+3\alpha^2)
-\alpha^2r_+^2\cos^2\theta+a^2\lambda_+^2>0,
\qquad \lambda_+=\frac{2r_+}{r_+-r_-}.
\label{eq:generalhorizonradicandlimit}
\end{equation}
Thereby, the field is real and has a smooth extension to a collar on both sides of the nonextremal horizon.


\subsection{Physical reconstruction and verification of the vacuum equations}

For the simple positive norm field, the physical metric is
\begin{equation}
\begin{split}
\mathrm ds_g^2={}&\frac{1}{C}\mathrm ds_h^2
+\frac{\xi_2 A}{C\,\Sigma}\left[
\Delta(\psi\,\mathrm d\theta+F_r\zeta)^2
+(\psi\,\mathrm dr-F_\theta\zeta)^2
+D(\mathrm dF)^2\right].
\end{split}
\label{eq:spatialphysicalmetric}
\end{equation}
Despite the inherited label, this metric includes temporal as well as spatial corrections. The auxiliary norm~\eqref{eq:generalhorizonnorm} ensures
\begin{equation}
g^{\mu\alpha}g^{\nu\beta}b_{\mu\nu}b_{\alpha\beta}=b^2,
\qquad \sqrt{-g}=\frac{\Sigma\sin\theta}{AC}.
\label{eq:generalhorizonphysicalnorm}
\end{equation}
The algebraic relation gives
$\sqrt{-g}K^{\mu\nu}=\sqrt{-h}h^{\mu\nu}$, so the independent connection is the Kerr connection. Since $R_{\mu\nu}(h)=0$, the closed field with $V=V'=0$ satisfies the metric and Kalb--Ramond equations. Positive $A$ and $C$ preserve the Lorentzian signature. In addition, the complete configuration is genuinely an exact vacuum solution.


\subsection{Smoothness and the physical Killing horizon}

In ingoing Kerr coordinates,
\begin{equation}
\mathrm dv_h=\mathrm dt+\frac{q}{\Delta}\,\mathrm dr,
\qquad
\mathrm d\phi_h=\mathrm d\phi+\frac{a}{\Delta}\,\mathrm dr,
\label{eq:generalhorizoningco}
\end{equation}
the corotating one form becomes
\begin{equation}
\zeta=\mathrm d\phi_h-\Omega_H\,\mathrm dv_h
+\Omega_H\lambda_r\,\mathrm dr.
\label{eq:generalhorizonregularzeta}
\end{equation}
Since $r_+>r_-$, every coefficient is finite at $r_+$. Furthermore, the two form reads
\begin{equation}
\boldsymbol b=(\psi-\Omega_H\lambda_rF_\theta)
\,\mathrm dr\wedge\mathrm d\theta
+\mathrm dF\wedge(\mathrm d\phi_h-\Omega_H\,\mathrm dv_h),
\label{eq:generalhorizonregularfield}
\end{equation}
Writing $\widehat\psi=\psi-\Omega_H\lambda_rF_\theta$, the same tensor in the ingoing chart has explicitly the component matrix
\begin{equation*}
\bigl[b_{\mu\nu}\bigr]_{(v_h,r,\theta,\phi_h)}=
\begin{pmatrix}
0 & \Omega_H F_r & \Omega_H F_\theta & 0\\
-\Omega_H F_r & 0 & \widehat\psi & F_r\\
-\Omega_H F_\theta & -\widehat\psi & 0 & F_\theta\\
0 & -F_r & -F_\theta & 0
\end{pmatrix},
\end{equation*}
with smooth coefficients. At either rotation axis,
$\psi=O(\sin\theta)$, $F_r=O(\sin^2\theta)$, and
$F_\theta=O(\sin\theta)$. These powers combine with the angular differentials into smooth tensors in local Cartesian coordinates, with the usual $2\pi$ period of $\phi_h$. The auxiliary metric, independent connection, and physical metric are smooth in this same extension.

Now, let us regard
\begin{equation}
\chi=\partial_t+\Omega_H\partial_\phi.
\label{eq:generalhorizonchi}
\end{equation}
Both metrics and the field are invariant under $\chi$. The defining property of the setting is
\begin{equation}
\iota_\chi\boldsymbol b=0,
\qquad
 g_{\mu\nu}\chi^\nu=\frac{1}{C}h_{\mu\nu}\chi^\nu,
\qquad g(\chi,\chi)=\frac{1}{C}h(\chi,\chi).
\label{eq:generalhorizonpreservation}
\end{equation}
At $r_+$, the last two relations imply that $\chi$ is null and normal to $r=r_+$ for the physical metric. In this sense, $r=r_+$ is a regular physical Killing horizon for unequal $A$ and $C$. In the ingoing chart,
\begin{equation}
\left.g(\chi,\cdot)\right|_{r_+}
=\frac{\Sigma_+}{Cq_+}\,\mathrm dr.
\label{eq:generalhorizononenormal}
\end{equation}
With $\chi$ future directed, every future directed causal vector $X$ on the horizon obeys
$g(\chi,X)\leq0$, and $X(r)\leq0$. In other words, this is the local one way property of the future horizon. It does not depend on setting the deformation factors equal.


\subsection{Exterior causal conditions and the asymptotic end}

A regular Killing horizon is a local statement. Identifying it as an event horizon also requires specifying the exterior end and showing that exterior points can escape to it. Based on this, the following inequalities give a sufficient exterior causal domain. By defining
\begin{equation}
\epsilon=\frac{C}{A}-1,\qquad
\mathcal A=\frac{DU}{k^2\Sigma^2},\qquad
\mathcal P=\frac{U}{k^2zW}=\frac{L_F}{Wd^3},\qquad
\mathcal Z=\mathcal P-\mathcal A\geq0.
\label{eq:generalhorizoncausalfunctions}
\end{equation}
The symbol $\mathcal P$ here denotes an exterior geometric function. The relevant exact identities are
\begin{align}
g_{\phi\phi}&=\frac{h_{\phi\phi}}{C}(1+\epsilon\mathcal P),
\qquad 0<\mathcal P\leq1,
\label{eq:generalhorizonaxialcausality}\\
g_{tt}g_{\phi\phi}-g_{t\phi}^2
&=-\frac{\Delta z}{C^2}(1+\epsilon\mathcal A),
\label{eq:generalhorizonstationaryblock}\\
g^{tt}&=A h^{tt}[1+\epsilon(1+\mathcal Z)],
\label{eq:generalhorizontemporal}\\
g^{rr}&=\frac{A\Delta}{\Sigma}
\left[1+\epsilon\frac{DF_\theta^2}{k^2\Sigma^2}\right].
\label{eq:generalhorizonradialinverse}
\end{align}
Because $\epsilon>-1$, the axial orbits remain spacelike. If
\begin{equation}
1+\epsilon\mathcal A>0,\qquad
1+\epsilon(1+\mathcal Z)>0
\qquad\text{throughout }r>r_+,
\label{eq:generalhorizoncausalconditions}
\end{equation}
the Killing block is Lorentzian, $t$ is a global time function on the exterior, and $g^{rr}>0$. A timelike stationary direction exists at every exterior point. Adding a sufficiently small outward radial component produces a future directed timelike path to arbitrarily large radius; an exterior point can also move away from the axes before escaping. Together with Eq.~\eqref{eq:generalhorizononenormal}, these properties identify the horizon as the escape boundary relative to an exterior asymptotic end admitting the corresponding future null boundary.

The sufficient domain is nonempty for every fixed nonextremal spin and includes unequal deformation factors. A conservative explicit bound is obtained from
\begin{equation}
\mathcal M=\frac{64a^2r_+}{27q_+(r_+-r_-)},\qquad
-\mathcal M\leq\mathcal A\leq1,\qquad
0\leq\mathcal Z\leq1+\mathcal M.
\label{eq:generalhorizonbounds}
\end{equation}
For example,
\begin{equation}
\left|\frac{C}{A}-1\right|<\frac{1}{2+\mathcal M}
\label{eq:generalhorizonsufficientdomain}
\end{equation}
ensures the inequalities~\eqref{eq:generalhorizoncausalconditions}. This is a sufficient bound, not a necessary restriction on the entire solution family. It becomes conservative near extremality. To justify the lower bound in Eq.~\eqref{eq:generalhorizonbounds}, use $\Sigma\geq q/2$, $L_F\leq2q^2$, $\lambda_r\leq2r_+/(r_+-r_-)$, and
$y/(1+y)^3\leq4/27$ for $y\geq0$.

The asymptotic qualification is essential. For fixed $0<\theta<\pi$, the leading metric is
\begin{equation}
\mathrm ds_g^2\sim-\frac{\mathrm dt^2}{C}
+\frac{\mathrm dr^2+r^2\mathrm d\theta^2}{A}
+\frac{r^2\sin^2\theta}{C}\,\mathrm d\phi^2.
\label{eq:generalhorizonasymptoticmetric}
\end{equation}
With
\begin{equation}
T=\frac{t}{\sqrt C},\qquad R=\frac{r}{\sqrt A},\qquad
\Phi=\sqrt{\frac AC}\,\phi,
\label{eq:canonicaltransformation}
\end{equation}
this limit is locally flat away from the axes, but $\Phi$ has period $2\pi\sqrt{A/C}$. The polar limit is nonuniform: regions with $\sin\theta=O(1/r)$ retain a distinct finite field profile. An important point needs to be mentioned: smoothness of the axes at every finite $r$ does not establish a smooth, standard asymptotically flat conformal completion at the polar endpoints of null infinity. The exact result established here is a regular physical Killing horizon and a sufficient causal exterior with an asymptotically locally flat end away from the axes.

The unequal factor metric is generally not Kerr, and its Hamilton--Jacobi separability is not established by this construction. A geodesic analysis must use the reconstructed metric itself. Equality of the deformation factors can be presented as a simplifying special limit; it is not a condition imposed by the local vacuum equations or by smoothness of the physical horizon in this enlarged configuration.


\subsection{Alternative vacuum configurations and the horizon criterion}
\label{subsec:alternativeKRconfigurations}

The auxiliary construction also makes it possible to explore different orientations of the Kalb--Ramond vacuum. A simple, closed two form with the prescribed norm determines an exact physical geometry through the same algebraic relation. The freedom in choosing its components gives a family of stationary configurations. A black hole interpretation adds a geometric requirement to this freedom: the physical horizon and the independent fields must admit a common smooth extension, together with the exterior causal and asymptotic structure discussed above. The following examples clarify how these requirements select the admissible horizon configurations.

A useful first choice is the purely meridional field
\begin{equation}
\boldsymbol b_{\rm mer}
=\frac{k\Sigma}{\sqrt\Delta}\,
\mathrm dr\wedge\mathrm d\theta.
\label{eq:alternativeKRmeridional}
\end{equation}
Its explicit component matrix is
\begin{equation*}
\bigl[(b_{\rm mer})_{\mu\nu}\bigr]_{(t,r,\theta,\phi)}
=\frac{k\Sigma}{\sqrt\Delta}
\begin{pmatrix}
0 & 0 & 0 & 0\\
0 & 0 & 1 & 0\\
0 & -1 & 0 & 0\\
0 & 0 & 0 & 0
\end{pmatrix}.
\end{equation*}
It is simple and closed, and its auxiliary norm is $2k^2=b^2/A^2$. The physical line element becomes
\begin{equation}
\begin{split}
\mathrm ds_{\rm mer}^2={}&\frac{1}{C}
\left\{-\frac{\Delta}{\Sigma}
(\mathrm dt-a\sin^2\theta\,\mathrm d\phi)^2
+\frac{\sin^2\theta}{\Sigma}
(q\,\mathrm d\phi-a\,\mathrm dt)^2\right\} +\frac{\Sigma}{A}
\left(\frac{\mathrm dr^2}{\Delta}+\mathrm d\theta^2\right).
\end{split}
\label{eq:alternativeKRmeridionalmetric}
\end{equation}
This expression displays directly the rescaling of the meridional plane and its orthogonal complement. Its metric horizon can be described in a regular chart, but the complete tensor configuration has a further requirement. Since $r$ and $\theta$ are unchanged by the ingoing transformation, the coefficient of $\mathrm dr\wedge\mathrm d\theta$ in Eq.~\eqref{eq:alternativeKRmeridional} still diverges as $\Delta^{-1/2}$ at $r_+$. It also approaches a nonzero value at either axis, where a smooth coefficient multiplying this angular one form must vanish appropriately. Taking all these features into account, this field shows an exact exterior vacuum and an algebraic example, while the corotating profile developed above supplies the additional components needed for a smooth horizon and axis extension of the independent fields.

On the other hand, a broader comparison retains the same flux function $F$ and introduces a constant angular parameter $\omega$:
\begin{equation}
\boldsymbol b_\omega
=\psi_\omega\,\mathrm dr\wedge\mathrm d\theta
+\mathrm dF\wedge\zeta_\omega,
\qquad
\zeta_\omega=\mathrm d\phi-\omega\,\mathrm dt.
\label{eq:alternativeKRomegaansatz}
\end{equation}
The corresponding matrix makes the temporal and spatial components explicit:
\begin{equation*}
\bigl[(b_\omega)_{\mu\nu}\bigr]_{(t,r,\theta,\phi)}=
\begin{pmatrix}
0 & \omega F_r & \omega F_\theta & 0\\
-\omega F_r & 0 & \psi_\omega & F_r\\
-\omega F_\theta & -\psi_\omega & 0 & F_\theta\\
0 & -F_r & -F_\theta & 0
\end{pmatrix}.
\end{equation*}
The nonzero independent components are
\begin{equation}
\begin{gathered}
(b_\omega)_{r\theta}=\psi_\omega,\qquad
(b_\omega)_{r\phi}=F_r,\qquad
(b_\omega)_{\theta\phi}=F_\theta,\\
(b_\omega)_{tr}=\omega F_r,\qquad
(b_\omega)_{t\theta}=\omega F_\theta.
\end{gathered}
\label{eq:alternativeKRomegacomponents}
\end{equation}
For every constant $\omega$, simplicity and closure follow as before. Define
\begin{equation}
D_\omega=\frac{(1-a\omega z)^2}{z}
-\frac{(a-\omega q)^2}{\Delta},\qquad
\mathcal S_\omega=\Delta\Sigma^2
-\frac{\Delta D_\omega U}{k^2}.
\label{eq:alternativeKRomegaradicand}
\end{equation}
The fixed norm condition gives two exterior scenarios,
\begin{equation}
\psi_{\omega,\pm}=\pm\frac{k\sqrt{\mathcal S_\omega}}{\Delta},
\qquad
\Delta\psi_\omega^2+D_\omega U=k^2\Sigma^2.
\label{eq:alternativeKRomegabranches}
\end{equation}
Both cases can be written explicitly as
\begin{equation*}
\bigl[(b_{\omega,\pm})_{\mu\nu}\bigr]_{(t,r,\theta,\phi)}=
\begin{pmatrix}
0 & \omega F_r & \omega F_\theta & 0\\
-\omega F_r & 0 & \displaystyle\pm\frac{k\sqrt{\mathcal S_\omega}}{\Delta} & F_r\\[0.6em]
-\omega F_\theta & \displaystyle\mp\frac{k\sqrt{\mathcal S_\omega}}{\Delta} & 0 & F_\theta\\[0.6em]
0 & -F_r & -F_\theta & 0
\end{pmatrix}.
\end{equation*}
These profiles are real on the exterior for the flux function used here. Indeed,
$D_\omega\leq\Sigma^2/(zW)$ for any real $\omega$, and the previously established inequality $Wd^3-L_F>0$ ensures $\mathcal S_\omega>0$ for $r>r_+$ away from the axes. Their algebraic reconstruction is
\begin{equation}
\begin{split}
\mathrm ds_\omega^2={}&\frac{1}{C}\mathrm ds_h^2
+\frac{\xi_2 A}{C\Sigma}\left[
\Delta(\psi_\omega\,\mathrm d\theta+F_r\zeta_\omega)^2
+(\psi_\omega\,\mathrm dr-F_\theta\zeta_\omega)^2
+D_\omega(\mathrm dF)^2\right].
\end{split}
\label{eq:alternativeKRomegametric}
\end{equation}
In particular, $\omega=0$ gives a purely spatial configuration with radial, polar, and azimuthal support. Its two settings have the matrices
\begin{equation*}
\bigl[(b_{0,\pm})_{\mu\nu}\bigr]_{(t,r,\theta,\phi)}=
\begin{pmatrix}
0 & 0 & 0 & 0\\
0 & 0 & \displaystyle\pm\frac{k\sqrt{\mathcal S_0}}{\Delta} & F_r\\[0.6em]
0 & \displaystyle\mp\frac{k\sqrt{\mathcal S_0}}{\Delta} & 0 & F_\theta\\[0.6em]
0 & -F_r & -F_\theta & 0
\end{pmatrix},
\qquad \mathcal S_0=\mathcal S_\omega\big|_{\omega=0}.
\end{equation*}
Nonzero $\omega$ introduces temporal components and changes the physical Killing block. The two radial configurations also change mixed components, for example
\begin{equation}
(g_\omega)_{r\phi}
=-\frac{\xi_2 A\psi_\omega F_\theta}{C\Sigma},\qquad
(g_\omega)_{rt}
=\frac{\xi_2 A\omega\psi_\omega F_\theta}{C\Sigma}.
\label{eq:alternativeKRmixedcomponents}
\end{equation}
Then, the norm and closure conditions allow different physical metric expressions while retaining the same Ricci flat auxiliary geometry.

The future extension distinguishes the two configurations when $\omega\ne\Omega_H$. In ingoing Kerr coordinates, the meridional coefficient of the complete two form is
\begin{equation}
\widehat\psi_\omega
=\psi_\omega+\frac{a-\omega q}{\Delta}F_\theta.
\label{eq:alternativeKRingoingcoefficient}
\end{equation}
The ingoing components of either setting are accordingly
\begin{equation*}
\bigl[(b_\omega)_{\mu\nu}\bigr]_{(v_h,r,\theta,\phi_h)}=
\begin{pmatrix}
0 & \omega F_r & \omega F_\theta & 0\\
-\omega F_r & 0 & \widehat\psi_\omega & F_r\\
-\omega F_\theta & -\widehat\psi_\omega & 0 & F_\theta\\
0 & -F_r & -F_\theta & 0
\end{pmatrix}.
\end{equation*}
Let $B_\omega=a-\omega q_+\ne0$. Since
$F_\theta|_{r_+}=kq_+\sin\theta$,
$\mathcal S_\omega|_{r_+}=B_\omega^2q_+^2\sin^2\theta$.
Choosing the sign in Eq.~\eqref{eq:alternativeKRomegabranches} to be $-\operatorname{sgn}(B_\omega)$ cancels the leading pole in Eq.~\eqref{eq:alternativeKRingoingcoefficient} and gives a smooth future extension. The opposite sign gives
\begin{equation}
\widehat\psi_\omega
=\frac{2B_\omega F_\theta|_{r_+}}{\Delta}+O(1).
\label{eq:alternativeKRoppositebranch}
\end{equation}
That segment is instead regular in the outgoing chart adapted to the past auxiliary horizon. This distinction identifies the appropriate extension for the intended physical interpretation. Reversing the complete two form, $k\mapsto-k$, changes its orientation without changing the reconstructed metric or the choice of future extension.

For the future regular part, the physical null condition further distinguishes the corotating configuration. Contraction with $\chi=\partial_t+\Omega_H\partial_\phi$ gives
\begin{equation}
\iota_\chi\boldsymbol b_\omega
=(\omega-\Omega_H)\,\mathrm dF.
\label{eq:alternativeKRgeneratorcontraction}
\end{equation}
The inverse metric satisfies
\begin{equation}
\left.g_\omega^{rr}\right|_{r_+}
=\frac{(A-C)(a-\omega q_+)^2q_+^2\sin^2\theta}
{(r_+^2+a^2\cos^2\theta)^3}.
\label{eq:alternativeKRhorizoninverse}
\end{equation}
Accordingly, for unequal factors the auxiliary surface $r=r_+$ is a physical null hypersurface throughout its angular range precisely when $\omega=\Omega_H$ within this family. For the purely spatial member $\omega=0$, its normal instead has the nonzero norm obtained by replacing $(a-\omega q_+)^2$ with $a^2$ in Eq.~\eqref{eq:alternativeKRhorizoninverse}. A physical horizon at a different position remains a separate possibility. A stationary axisymmetric candidate $r=\rho_\omega(\theta)$ would satisfy
\begin{equation}
g_\omega^{rr}-2\rho_\omega' g_\omega^{r\theta}
+(\rho_\omega')^2g_\omega^{\theta\theta}=0
\qquad\text{at }r=\rho_\omega(\theta),
\label{eq:alternativeKRnullcandidate}
\end{equation}
and its interpretation as an event horizon would additionally require the common extension and global escape analysis.

In other words, all these examples make the role of the vacuum orientation explicit. The meridional field gives a simple exact exterior geometry, the angular family provides further stationary solutions, and its radial configurations distinguish future and past extensions. The corotating member aligns the tensor kernel with the horizon generator and retains the regular physical Killing horizon without setting $A=C$. In this manner, the construction selects a definite horizon interpretation through properties of the complete configuration. Profiles with a demonstrated tensor obstruction do not provide that smooth extension; profiles whose physical event horizon has not yet been located remain exact vacuum candidates for further analysis. This distinction preserves the scope of the local solutions while making the requirements for their black hole interpretation precise.


\section{Geodesic motion in the exterior geometry}
\label{sec:geodesics}

We consider neutral, structureless particles minimally coupled to the physical metric reconstructed from the corotating Kalb--Ramond profile of Sec.~\ref{sec:generalhorizon}. Their trajectories are geodesics of the Levi--Civita connection of $g_{\mu\nu}$. The independent connection, which is compatible with the auxiliary Kerr metric, does not enter the matter equation of motion. Minimally coupled electromagnetic radiation follows null geodesics in the geometric optics limit. We work in an exterior parameter domain satisfying the causal conditions established in the preceding section.

The analysis starts from the physical metric itself. We use the asymptotically normalized time $T=t/\sqrt C$ and retain the physical azimuth $\phi$ with period $2\pi$. The radial coordinate $r$ remains the auxiliary Boyer--Lindquist coordinate.


\subsection{Conserved quantities and the reduced Hamiltonian}

Let $\lambda$ be proper time for a massive particle and an affine parameter for a null trajectory. With a dot denoting $\mathrm d/\mathrm d\lambda$, the normalization and momenta are
\begin{equation}
g_{\mu\nu}\dot x^\mu\dot x^\nu=-\eta,\qquad
p_\mu=g_{\mu\nu}\dot x^\nu,\qquad
\eta=\begin{cases}1,&\text{massive particles},\\0,&\text{null trajectories}.
\end{cases}
\label{eq:geodesicnormalization}
\end{equation}
Stationarity and axial symmetry give
\begin{equation}
\mathcal E=-p_T,\qquad L=p_\phi.
\label{eq:orbitalconstants}
\end{equation}
These constants are understood per unit rest mass for massive particles. If $E=-p_t$, then $\mathcal E=\sqrt C\,E$. The affine scale of photon constants is arbitrary, while their ratios are invariant under a constant rescaling of the affine parameter.

Let lower case Latin indices $i,j$ run over $r,\theta$ and capital Latin indices $A,B$ run over $T,\phi$. Set $\ell_A=(-\mathcal E,L)$. The exact reduced Hamiltonian is
\begin{equation}
\mathcal H=\frac12\left(
 g^{ij}p_i p_j+2g^{iA}p_i\ell_A
 +g^{AB}\ell_A\ell_B\right)=-\frac{\eta}{2}.
\label{eq:generalgeodesichamiltonian}
\end{equation}
Its first order equations are
\begin{align}
\dot x^i&=g^{ij}p_j+g^{iA}\ell_A,\\
\dot p_i&=-\frac12\partial_i g^{\mu\nu}p_\mu p_\nu,\\
\dot x^A&=g^{Ai}p_i+g^{AB}\ell_B.
\label{eq:generalgeodesicfirstintegrals}
\end{align}
This system specifies arbitrary timelike and null trajectories without a separability assumption. The Hamilton--Jacobi ansatz
$S=\eta\lambda/2-\mathcal E T+L\phi+S_{r\theta}(r,\theta)$ is admissible, but no decomposition $S_{r\theta}=S_r+S_\theta$ or additional constant $Q$ is assumed. The Kerr separated equations are recovered only in the equal factor limit of this particular reconstruction, as we should naturally expect.


\subsection{Equatorial motion and effective potentials}

The reconstructed metric is invariant under $\theta\mapsto\pi-\theta$. In this sense, equatorial initial data are preserved by the metric geodesic equations, even though the vacuum tensor itself need not be invariant under this reflection. At $\theta=\pi/2$, the components with one $\theta$ index vanish, and the remaining metric can be written
\begin{equation}
\mathrm ds^2=G_{AB}(r)\,\mathrm dx^A\mathrm dx^B
+2w_A(r)\,\mathrm dx^A\mathrm dr
+g_{rr}(r)\,\mathrm dr^2
+g_{\theta\theta}(r)\,\mathrm d\theta^2,
\label{eq:generalgeodesicequatorialmetric}
\end{equation}
where $G_{AB}=g_{AB}$ and $w_A=g_{rA}$. Introduce
\begin{equation}
\mathcal D=g_{T\phi}^2-g_{TT}g_{\phi\phi}>0,
\qquad
\gamma_{rr}=g_{rr}-w_A(G^{-1})^{AB}w_B>0.
\label{eq:generalgeodesicSchur}
\end{equation}
On this plane $g^{rr}=1/\gamma_{rr}$. The cyclic first integrals and normalization give
\begin{align}
\dot x^A&=(G^{-1})^{AB}(\ell_B-w_B\dot r),
\label{eq:generalgeodesiccyclic}\\
\dot r^2&=g^{rr}\left[
\frac{g_{\phi\phi}\mathcal E^2
+2g_{T\phi}\mathcal E L+g_{TT}L^2}{\mathcal D}
-\eta\right]\equiv\mathcal F(r;\mathcal E,L).
\label{eq:generalgeodesicradial}
\end{align}
The mixed radial terms enter through the Schur complement and the cyclic first integrals. Notice that they cannot simply be omitted when computing radial frequencies or azimuthal changes.

Factoring the radial equation with respect to the energy yields
\begin{align}
\mathcal V_\pm(r;L)&=
\frac{-g_{T\phi}L\pm
\sqrt{\mathcal D(L^2+\eta g_{\phi\phi})}}
{g_{\phi\phi}},
\label{eq:KReffectivepotentials}\\
\dot r^2&=\frac{g^{rr}g_{\phi\phi}}{\mathcal D}
(\mathcal E-\mathcal V_+)(\mathcal E-\mathcal V_-).
\label{eq:radialfactorization}
\end{align}
The prefactor is positive in the admissible exterior. At a radial turning point,
$\dot T=(g_{\phi\phi}\mathcal E+g_{T\phi}L)/\mathcal D$, so the nonzero future directed turning root is $\mathcal V_+$. The coordinate radial acceleration is
\begin{equation}
\ddot r=\frac12\partial_r\mathcal F,
\label{eq:radialacceleration}
\end{equation}
with the conserved quantities held fixed; the same expression follows from the geodesic equation by continuity at a turning point.

For explicit evaluation, let $q=r^2+a^2$, $q_+=r_+^2+a^2$, $\Omega_H=a/q_+$, $\epsilon=C/A-1$, and
$K(r)=q_+^3/(qr^2)$. The equatorial Killing block of the new profile is
\begin{align}
g_{TT}&=-1+\frac{2m}{r}+\epsilon\Omega_H^2K,\\
g_{T\phi}&=\frac{-2ma/r-\epsilon\Omega_H K}{\sqrt C},\\
g_{\phi\phi}&=\frac{q+2ma^2/r+\epsilon K}{C}.
\label{eq:generalgeodesicexplicitblock}
\end{align}
The radial and polar components, including the mixed terms, follow from the full reconstructed metric in Sec.~\ref{sec:generalhorizon}. Eqs.~\eqref{eq:generalgeodesicradial}--\eqref{eq:generalgeodesicexplicitblock} are exact for this configuration.


\subsection{Circular orbits and radial stability}

For a circular equatorial timelike orbit, $\mathcal F=\partial_r\mathcal F=0$. Its angular velocity $\Omega=\mathrm d\phi/\mathrm dT$ also follows directly from the radial geodesic equation:
\begin{equation}
\partial_r g_{TT}+2\Omega\partial_r g_{T\phi}
+\Omega^2\partial_r g_{\phi\phi}=0.
\label{eq:generalgeodesiccircularcondition}
\end{equation}
Where $\partial_r g_{\phi\phi}\ne0$, the two roots are
\begin{equation}
\Omega_\pm=\frac{-\partial_r g_{T\phi}
\pm\sqrt{(\partial_r g_{T\phi})^2
-(\partial_r g_{TT})(\partial_r g_{\phi\phi})}}
{\partial_r g_{\phi\phi}}.
\label{eq:circularangularvelocity}
\end{equation}
A real root describes a timelike circular orbit only when
$\mathcal N=-g_{TT}-2g_{T\phi}\Omega-g_{\phi\phi}\Omega^2>0$.
Its conserved quantities are
\begin{align}
\mathcal E_\pm&=-\frac{g_{TT}+g_{T\phi}\Omega_\pm}
{\sqrt{\mathcal N_\pm}},
\label{eq:circularenergy}\\
L_\pm&=\frac{g_{T\phi}+g_{\phi\phi}\Omega_\pm}
{\sqrt{\mathcal N_\pm}}.
\label{eq:circularangularmomentum}
\end{align}
The prograde or retrograde character is determined from the corresponding angular velocity and the orientation of the asymptotic frame.

At a circular radius $r_c$, linearization at fixed $\mathcal E_c,L_c$ gives
\begin{equation}
\omega_{r,T}^2=-\frac{\partial_r^2\mathcal F|_c}
{2\dot T_c^{\,2}}.
\label{eq:radialepicyclicfrequency}
\end{equation}
Radial stability requires $\partial_r^2\mathcal F|_c<0$. Marginal stability satisfies
\begin{equation}
\mathcal F=\partial_r\mathcal F
=\partial_r^2\mathcal F=0.
\label{eq:KRISCO}
\end{equation}
The exterior configuration connected to stable large radius orbits defines the innermost radially stable circular orbit. A full stability analysis also checks polar perturbations. Also, we have to notice that the fixed Kerr polynomial for its radius is not valid for unequal factors in this new metric.

For a circular equatorial photon orbit, set $\eta=0$ and define $b_{\rm imp}=L/\mathcal E$. Its radius and impact parameter are determined by
\begin{equation}
\mathcal F(r;1,b_{\rm imp})=0,\qquad
\partial_r\mathcal F(r;1,b_{\rm imp})=0.
\label{eq:generalgeodesicphotonconditions}
\end{equation}
In a similar manner, let us solve
$g_{TT}+2g_{T\phi}\Omega+g_{\phi\phi}\Omega^2=0$
together with Eq.~\eqref{eq:generalgeodesiccircularcondition}.
The second radial derivative determines radial instability. 


\subsection{Perihelion advance}

For a slightly eccentric, radially stable equatorial orbit, the excess azimuthal angle in the direction of motion is
\begin{equation}
\delta\phi=2\pi\left(\frac{|\Omega|}{\omega_{r,T}}-1\right).
\label{eq:nearCircularAdvance}
\end{equation}
Both frequencies are evaluated in the same normalized time, as we did before.

For an eccentric bound orbit between $r_{\rm peri}$ and $r_{\rm apo}$, define
\begin{equation}
\dot\phi_0=-\frac{g_{T\phi}\mathcal E+g_{TT}L}{\mathcal D}.
\label{eq:generalgeodesicphidotzero}
\end{equation}
The radial velocity term in Eq.~\eqref{eq:generalgeodesiccyclic} integrates to zero over an outward and inward radial cycle. The signed azimuthal change is 
\begin{equation}
\Delta\phi=2\int_{r_{\rm peri}}^{r_{\rm apo}}
\frac{\dot\phi_0(r)}{\sqrt{\mathcal F(r;\mathcal E,L)}}\,\mathrm dr.
\label{eq:generalgeodesicadvanceintegral}
\end{equation}
For an orbit maintaining a definite azimuthal orientation, its advance in that direction is $s\Delta\phi-2\pi$, where $s=\pm1$ denotes that orientation.

Far from the hole and away from the polar asymptotic region, the leading metric is the asymptotically locally flat geometry described in Sec.~\ref{sec:generalhorizon}. A weak field expansion must include that angular normalization as well as the metric corrections. The previous exact Kerr Binet equation and the claim that all local orbital observables maintain Kerr values are not used here. It is important to mention that quantitative weak field coefficients and observational bounds must be derived with a common choice of asymptotic end, angular convention, and dynamical mass setting instead.


\subsection{Representative timelike and null trajectories}
\label{subsec:numerical-geodesics}

\begin{figure}[htbp]
    \centering
    \includegraphics[width=0.99\linewidth]{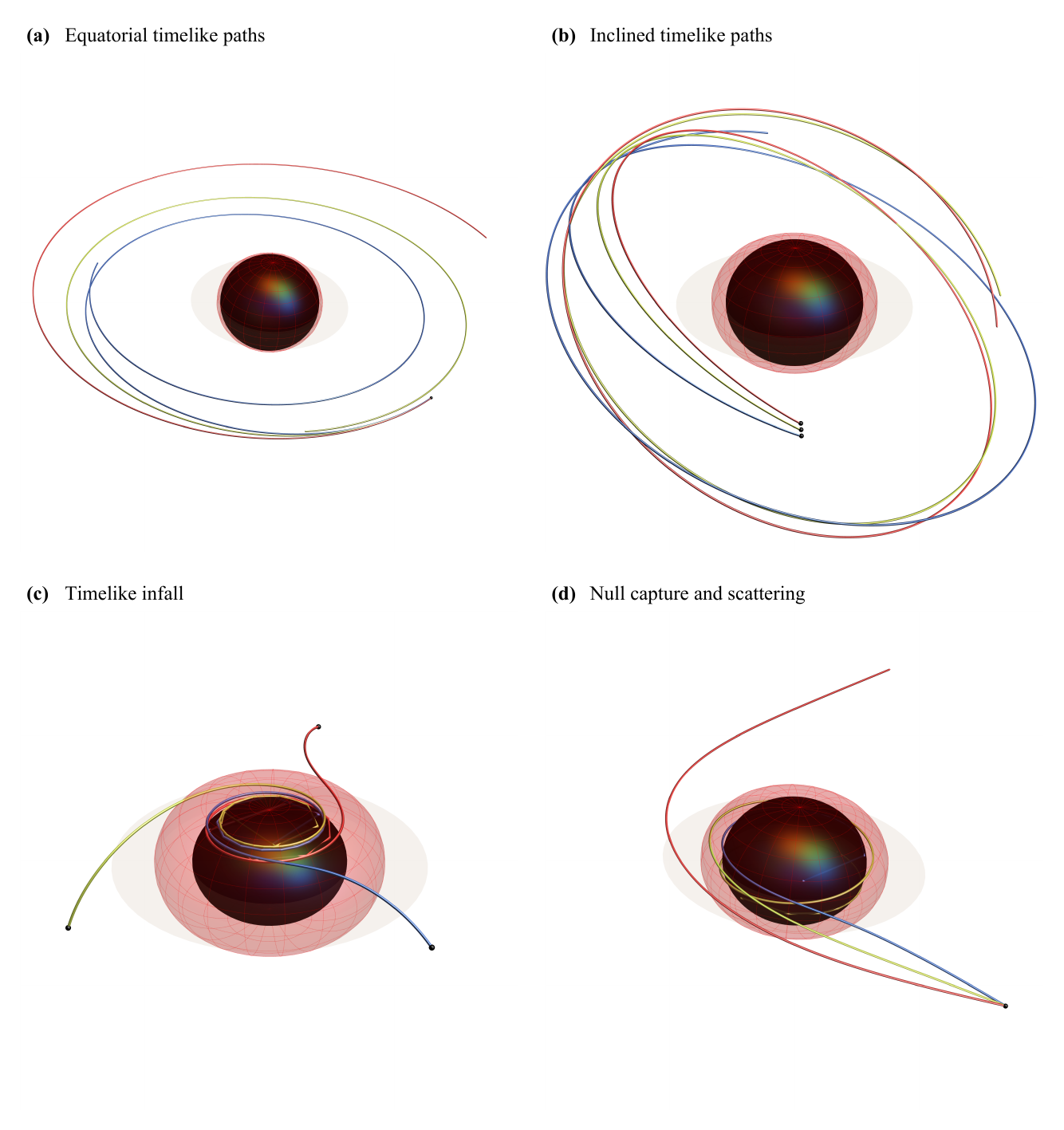}
    \caption{Representative geodesics in the rotating Palatini Kalb--Ramond geometry, with $A=1$ and $\epsilon=C/A-1$. (a) Equatorial timelike motion: $a/m=0.5$, $\epsilon=0.05$. (b) Inclined timelike motion: $a/m=0.8$, $\epsilon=0.05$.     (c) Timelike infall: $a/m=0.99$, $\epsilon=0.10$. (d) Null capture and outward scattering: $a/m=0.9$, $\epsilon=0.05$. The dark inner surface is the horizon; the translucent red surface is the stationary limit surface $g_{TT}=0$, bounding the ergoregion     together with the horizon. The gray disk marks the equatorial plane, and black dots indicate initial positions. Panels (a) and (b) extend to $\tau/m=120$. Captured trajectories terminate at     $r=r_++10^{-3}m$; the outward ray in (d) terminates at $r=10m$. The panels have different spatial scales and were made with \textit{Mathematica} software.}
    \label{fig:geodesic-configurations}
\end{figure}

In Fig.~\ref{fig:geodesic-configurations}, we illustrate representative timelike and null trajectories obtained by integrating the Hamiltonian equations for the physical metric. We use $A=1$ throughout and characterize the unequal factor deformation by $\epsilon=C/A-1$. The initial time component is fixed by the future-directed mass shell condition, $g_{\mu\nu}\dot{x}^{\mu}\dot{x}^{\nu}=-\eta$, with $\eta=1$ for timelike motion and $\eta=0$ for null motion. Dots denote derivatives with respect to proper time $\tau$ or the chosen null affine parameter $\lambda$, respectively.

In Fig.~\ref{fig:geodesic-configurations}(a), three equatorial timelike geodesics are shown for $a/m=0.5$ and $\epsilon=0.05$. They start at $r_0=8m$, $\theta_0=\pi/2$, and $\varphi_0=0$, with $\dot r_0=\dot\theta_0=0$. The blue, yellow green, and red curves have $m\dot\varphi_0=0.053$, $0.055$, and $0.058$, respectively. Over $0\leq\tau/m\leq120$, the blue trajectory develops an inward radial excursion, the yellow green trajectory remains nearly circular, and the red trajectory reaches larger radii. The sampled radial intervals are approximately $6.038\leq r/m\leq8$, $7.676\leq r/m\leq8$, and $8\leq r/m\leq10.755$, respectively. All three remain in the equatorial plane, with the reflection symmetry of the physical metric.

In Fig.~\ref{fig:geodesic-configurations}(b), nonzero initial polar velocities produce three dimensional timelike trajectories at $a/m=0.8$ and $\epsilon=0.05$. In the same color order, their initial radii are $r_0/m=(6,5.75,5.5)$, their polar velocities are $m\dot\theta_0=(0.035,0.050,0.060)$, and their azimuthal velocities are $m\dot\varphi_0=(0.085,0.085,0.090)$. Each begins at $\theta_0=\pi/2$ and $\varphi_0=0$, with $\dot r_0=0$. The different polar excursions and spatial orientations over $0\leq\tau/m\leq120$ illustrate the coupled radial, polar, and azimuthal evolution. These trajectories are calculated from the full Hamiltonian without assuming an additional separation constant.

In Fig.~\ref{fig:geodesic-configurations}(c), three inward timelike trajectories start at $r_0=4m$ with different polar angles, azimuths, and velocities for $a/m=0.99$ and $\epsilon=0.10$. All three approach the horizon while accumulating azimuthal winding. The integration terminates at $r=r_++10^{-3}m$, so the curves resolve the exterior approach to the horizon.

In Fig.~\ref{fig:geodesic-configurations}(d), three initially inward null geodesics are displayed for $a/m=0.9$ and $\epsilon=0.05$. They share $r_0=6m$, $\theta_0=\pi/2-0.2$, $\varphi_0=0$, $\dot r_0=-0.4$, and $m\dot\theta_0=0.01$, while $m\dot\varphi_0=0.02$, $0.04$, and $0.06$ for the blue, yellow green, and red rays. The blue ray reaches the horizon buffer; the yellow green ray accumulates approximately $2.28$ turns in the coordinate azimuth before reaching the same boundary. The red ray reaches a minimum radius $r_{\min}\simeq2.918m$, turns outward, and is followed to $r=10m$.

The spatial visualization uses $x+iy=\sqrt{r^2+a^2}\sin\theta\,e^{i\varphi}$ and $z=r\cos\theta$. It is a coordinate representation, and the panels use different spatial scales. The finite trajectory segments illustrate selected orbital outcomes.

\section{Photon regions, shadows, and observational constraints}
\label{sec:shadows}

We now examine the propagation of light in the physical geometry of Eq.~\eqref{eq:spatialphysicalmetric}. The corotating Kalb--Ramond configuration constructed in Sec.~\ref{sec:generalhorizon} gives the horizon and the exterior causal domain used throughout this analysis. We retain minimal electromagnetic coupling and negligible plasma refraction, so photons follow the null geodesics discussed in Sec.~\ref{sec:geodesics}. Their directions are measured in an orthonormal frame carried by the observer. This local configuration connects the reconstructed metric to the image on the celestial screen.

The critical curve separates photon histories that reach the exterior illumination from those that enter the absorbing region at the horizon. Its interpretation as a shadow uses this illumination and absorption prescription~\cite{Perlick2022}. The calculation can be formulated at finite observer radius, where the metric and its rotation axes have the smooth extension established in Sec.~\ref{sec:generalhorizon}. The asymptotic interpretation retains the qualification concerning the polar endpoints of the exterior end.


\subsection{Normalized coordinates and the physical deformation}

For an observer away from the rotation axes, the coordinates
\begin{equation}
T=\frac{t}{\sqrt C},\qquad R=\frac r{\sqrt A},\qquad
\Phi=\sqrt{\frac AC}\,\phi
\nonumber
\end{equation}
give the leading asymptotic line element
$-\mathrm dT^2+\mathrm dR^2+R^2(\mathrm d\theta^2+\sin^2\theta\,\mathrm d\Phi^2)$
at fixed $0<\theta<\pi$. We introduce
\begin{equation}
\mu=\frac m{\sqrt A},\qquad a_{\mathrm p}=\frac a{\sqrt A},\qquad
\chi=\frac{a_{\mathrm p}}\mu=\frac am,\qquad
\gamma=\frac CA=1+\epsilon,
\qquad \epsilon=\frac{\xi_2b^2}{2(1+\xi_1b^2)}.
\label{eq:shadownormalizedparameters}
\end{equation}
Here, $\mu$ is the length scale entering the normalized metric; a total mass at infinity additionally uses the global asymptotic structure. The angular identification is
$\Phi\sim\Phi+2\pi/\sqrt\gamma$. Accordingly, $\Phi$ is useful for local use, while the original azimuth with period $2\pi$ supplies regular coordinates around each finite radius axis.

The normalized coordinates also make the parameter dependence of the complete metric explicit. Let us write
\begin{equation}
\begin{gathered}
\Sigma_{\mathrm p}=R^2+a_{\mathrm p}^2\cos^2\theta,\qquad
\Delta_{\mathrm p}=R^2-2\mu R+a_{\mathrm p}^2,\qquad
q_{\mathrm p}=R^2+a_{\mathrm p}^2,\\
R_+=\mu+\sqrt{\mu^2-a_{\mathrm p}^2},\qquad
q_{\mathrm p+}=R_+^2+a_{\mathrm p}^2,\qquad
\Omega_{\mathrm p}=\frac{a_{\mathrm p}}{q_{\mathrm p+}},\qquad
\zeta_{\mathrm p}=\mathrm d\Phi-\Omega_{\mathrm p}\,\mathrm dT.
\end{gathered}
\label{eq:shadowphysicalfunctions}
\end{equation}
Define $\mathfrak F=F/(kA)$, $f_R=F_r/(k\sqrt A)$,
$f_\theta=F_\theta/(kA)$, $\Psi=\psi/(k\sqrt A)$, and $D_{\mathrm p}=D$ after the coordinate substitution. In this regard,
\begin{equation}
\mathfrak F=-\frac{q_{\mathrm p+}\cos\theta}{\sqrt d},\qquad
f_R=\frac{R\sin^2\theta\cos\theta}{d^{3/2}},\qquad
f_\theta=\frac{q_{\mathrm p}\sin\theta}{d^{3/2}},\qquad
d=1+\frac{q_{\mathrm p}-q_{\mathrm p+}}{q_{\mathrm p+}}\sin^2\theta.
\label{eq:shadownormalizedflux}
\end{equation}
The remaining functions are the expressions of Sec.~\ref{sec:generalhorizon} with $(r,m,a)$ replaced by $(R,\mu,a_{\mathrm p})$; in particular, $\mathcal B$ becomes $\mathcal B/A$. If $\mathrm ds_{\mathrm K}^2$ denotes the Kerr line element with parameters $(\mu,a_{\mathrm p})$, the exact physical metric is
\begin{equation}
\begin{split}
\mathrm ds_g^2=\mathrm ds_{\mathrm K}^2
+\frac{\epsilon}{\gamma}\Bigg\{&\frac1{\Sigma_{\mathrm p}}
\left[\Delta_{\mathrm p}(\Psi\,\mathrm d\theta+\sqrt\gamma f_R\zeta_{\mathrm p})^2
+(\Psi\,\mathrm dR-\sqrt\gamma f_\theta\zeta_{\mathrm p})^2
+D_{\mathrm p}(\mathrm d\mathfrak F)^2\right] -\Sigma_{\mathrm p}\left(\frac{\mathrm dR^2}{\Delta_{\mathrm p}}+\mathrm d\theta^2\right)\Bigg\}.
\end{split}
\label{eq:normalizedshadowmetric}
\end{equation}
In other words, the above expression maintains the temporal corrections and the mixed components of Eq.~\eqref{eq:spatialphysicalmetric}. After measuring lengths in units of $\mu$, the metric depends on $\chi$ and $\epsilon$. Normalization isolates the overall scale while preserving the physical deformation. As it is straight forward to verify, for $\epsilon=0$, Eq.~\eqref{eq:normalizedshadowmetric} turns out to be Kerr exactly.


\subsection{Photon dynamics and the trapped region}

Stationarity and axial symmetry give the conserved photon quantities
\begin{equation}
\mathcal E=-p_T,\qquad \mathcal L=p_\Phi=\sqrt\gamma\,p_\phi,
\qquad b_{\mathrm{imp}}=\frac{\mathcal L}{\mathcal E}.
\label{eq:shadowphotonconstants}
\end{equation}
We take $\mathcal E>0$ for radiation reaching the distant observer. The impact parameter has dimensions of length. With $i,j\in\{R,\theta\}$, $A,B\in\{T,\Phi\}$, and $\ell_A=(-\mathcal E,\mathcal L)$, the exact null Hamiltonian is
\begin{equation}
\mathcal H=\frac12\left(g^{ij}p_i p_j+2g^{iA}p_i\ell_A
+g^{AB}\ell_A\ell_B\right)=0.
\label{eq:shadowphysicalhamiltonian}
\end{equation}
Its equations of motion are
\begin{equation}
\dot x^\mu=g^{\mu\nu}p_\nu,\qquad
\dot p_i=-\frac12\partial_i g^{\mu\nu}p_\mu p_\nu.
\label{eq:shadowhamiltonequations}
\end{equation}
The dot denotes differentiation with respect to an affine parameter. The mixed terms enter both the positions and momenta and are retained in the numerical evolution.

The photon region is determined by trajectories that remain trapped in the exterior. Their radial and polar motion can be coupled, and the critical directions approach the trapped set separating capture from escape. The two conserved quantities in Eq.~\eqref{eq:shadowphotonconstants} give us the reduction required here. Thereby, the critical curve is obtained directly from Eq.~\eqref{eq:shadowhamiltonequations}, which allows the trapped trajectories to take the form selected by the geometry.

Equatorial circular photon trajectories show an important analytical characterization. Reflection symmetry preserves equatorial initial data. Let us write $G_{AB}=g_{AB}|_{\theta=\pi/2}$ in the normalized coordinates and set
\begin{equation}
K=\frac{q_{\mathrm p+}^3}{q_{\mathrm p}R^2}.
\end{equation}
Their exact Killing block is
\begin{align}
G_{TT}&=-1+\frac{2\mu}{R}+\epsilon\Omega_{\mathrm p}^2K,\nonumber\\
G_{T\Phi}&=-\frac{2\mu a_{\mathrm p}}R-\epsilon\Omega_{\mathrm p}K,\nonumber\\
G_{\Phi\Phi}&=q_{\mathrm p}+\frac{2\mu a_{\mathrm p}^2}R+\epsilon K.
\label{eq:shadowequatorialblock}
\end{align}
Now, let us define $\mathcal D=G_{T\Phi}^2-G_{TT}G_{\Phi\Phi}$ and
\begin{equation}
\mathcal V(R;b_{\mathrm{imp}})=G_{\Phi\Phi}
+2G_{T\Phi}b_{\mathrm{imp}}+G_{TT}b_{\mathrm{imp}}^2.
\label{eq:shadowequatorialpotential}
\end{equation}
The equatorial radial equation and the circular conditions are
\begin{equation}
\dot R^2=\frac{g^{RR}\mathcal E^2}{\mathcal D}\,\mathcal V,
\qquad \mathcal V=0,\qquad \partial_R\mathcal V=0,
\label{eq:shadowequatorialconditions}
\end{equation}
with the impact parameter fixed in the differentiation. The positive prefactor in the admissible exterior implies radial instability when $\partial_R^2\mathcal V>0$ at a circular trajectory. These conditions give the prograde and retrograde equatorial trajectories from the reconstructed metric. For an equatorial observer they supply checks on the intersections with $\beta=0$. In addition, the complete contour at arbitrary inclination follows from the full photon evolution.


\subsection{The observer's orthonormal screen}

Place the observer at $(R_{\mathrm o},\theta_{\mathrm o})$, with $0<\theta_{\mathrm o}<\pi$. All metric functions in the following frame are evaluated there. An observer at fixed radius and inclination with zero angular momentum has
\begin{equation}
\omega_{\mathrm o}=-\frac{g_{T\Phi}}{g_{\Phi\Phi}},\qquad
N_{\mathrm o}^2=\frac{g_{T\Phi}^2-g_{TT}g_{\Phi\Phi}}{g_{\Phi\Phi}},\qquad
u=\frac{\partial_T+\omega_{\mathrm o}\partial_\Phi}{N_{\mathrm o}}.
\label{eq:shadowphysicalobserver}
\end{equation}
The azimuthal unit vector is $e_{(\Phi)}=\partial_\Phi/\sqrt{g_{\Phi\Phi}}$. The radial and polar vectors are obtained by orthogonal projection:
\begin{align}
v_R&=\partial_R+g(u,\partial_R)u-g(e_{(\Phi)},\partial_R)e_{(\Phi)},
&e_{(R)}&=\frac{v_R}{\sqrt{g(v_R,v_R)}},\nonumber\\
v_\theta&=\partial_\theta+g(u,\partial_\theta)u
-g(e_{(\Phi)},\partial_\theta)e_{(\Phi)}
-g(e_{(R)},\partial_\theta)e_{(R)},
&e_{(\theta)}&=\frac{v_\theta}{\sqrt{g(v_\theta,v_\theta)}}.
\label{eq:shadowprojectedframe}
\end{align}
Together with $e_{(0)}=u$, these vectors satisfy $g(e_{(a)},e_{(b)})=\operatorname{diag}(-1,1,1,1)_{ab}$. This construction incorporates the mixed metric components and defines the radial vector to point outward.

The measured photon components are
\begin{equation}
p^{(0)}=-p_\mu u^\mu=\frac{\mathcal E-\omega_{\mathrm o}\mathcal L}{N_{\mathrm o}},
\qquad p^{(a)}=p_\mu e_{(a)}^\mu,
\qquad p^{(\Phi)}=\frac{\mathcal L}{\sqrt{g_{\Phi\Phi}}}.
\label{eq:shadowphysicalmomenta}
\end{equation}
For photons arriving with $p^{(R)}>0$, a plane tangent to the inward viewing direction has the linear coordinates
\begin{equation}
\alpha_{\mathrm o}=-R_{\mathrm o}\frac{p^{(\Phi)}}{p^{(R)}},\qquad
\beta_{\mathrm o}=R_{\mathrm o}\frac{p^{(\theta)}}{p^{(R)}}.
\label{eq:shadowfinitescreen}
\end{equation}
In an equivalent manner, each screen point specifies the future photon momentum at the observer through
\begin{equation}
\frac{p^\mu}{p^{(0)}}=u^\mu+
\frac{e_{(R)}^\mu-(\alpha_{\mathrm o}/R_{\mathrm o})e_{(\Phi)}^\mu
+(\beta_{\mathrm o}/R_{\mathrm o})e_{(\theta)}^\mu}
{\sqrt{1+(\alpha_{\mathrm o}^2+\beta_{\mathrm o}^2)/R_{\mathrm o}^2}}.
\label{eq:shadowlaunchdirection}
\end{equation}
The null condition is satisfied by construction. Evolving this momentum toward the past identifies the corresponding photon history.

For a distant observer at fixed nonpolar inclination, the convinient limits are
\begin{equation}
\alpha=-\frac{b_{\mathrm{imp}}}{\sin\theta_{\mathrm o}},\qquad
\beta=\lim_{R_{\mathrm o}\to\infty}
R_{\mathrm o}\frac{p^{(\theta)}}{p^{(0)}}.
\label{eq:shadowasymptoticscreen}
\end{equation}
The second coordinate is supplied by the evolved polar momentum. For $\epsilon\ne0$, the Hamiltonian construction retains its full dependence on the trajectory. The azimuthal period belongs to the global geometry, while the two screen axes have the same local orthonormal configuration.


\subsection{The critical curve and its geometric measures}

We construct the image by tracing Eq.~\eqref{eq:shadowhamiltonequations} from the observer to an absorbing surface just inside $R_+$ or to an illuminated outer boundary beyond the observer. Directions near the critical curve spend an increasing affine interval close to the trapped photon set. Resolving the transition between the two outcomes determines the contour $\mathcal C_{\mathrm{sh}}$. The integrations use Cartesian coordinates derived from the outgoing auxiliary Kerr chart, so rays crossing a finite radius axis have a regular continuation. Every direction is evaluated independently around the entire screen.

At a finite observer radius, the dimensionless contour depends on $R_{\mathrm o}/\mu$ as well as $(\chi,\theta_{\mathrm o},\epsilon)$. The distant observer scaling takes the form
\begin{equation}
\mathcal C_{\mathrm{sh}}=\mu\,\widehat{\mathcal C}_{\mathrm{sh}}
(\chi,\theta_{\mathrm o},\epsilon).
\label{eq:shadowphysicalscaling}
\end{equation}
By varying $A$ at fixed auxiliary mass can alter both the image scale through $\mu$ and the shape through $\epsilon$. At fixed $\mu$, $\chi$, inclination, and $\epsilon$, the normalized contour is the same. Holding the second coupling fixed while varying the first generally changes $\epsilon$ and give rises to a physical effect that survives the scale normalization.

The area and its associated radius are defined for the full closed curve by
\begin{equation}
\mathcal A_{\mathrm{sh}}=\frac12\left|\oint_{\mathcal C_{\mathrm{sh}}}
(\alpha\,\mathrm d\beta-\beta\,\mathrm d\alpha)\right|,
\qquad \rho_{\mathrm A}=\sqrt{\frac{\mathcal A_{\mathrm{sh}}}\pi},
\qquad d_{\mathrm A}=\frac{2\rho_{\mathrm A}}{\mathcal D_{\mathrm o}},
\label{eq:shadowgeneralarea}
\end{equation}
where $\mathcal D_{\mathrm o}$ is the physical distance used in the angular configuration. Horizontal and vertical extents and the area centroid give complementary measures of displacement and shape. Each normalized measure can depend on $\epsilon$ as well as rotation and inclination.

The Kerr limit supplies analytical checks. When $\epsilon=0$, the spherical photon impact parameters and the standard Kerr critical curve are recovered~\cite{Perlick2022}. With $\chi=0$ as well, the distant contour obeys
\begin{equation}
\alpha^2+\beta^2=27\mu^2.
\label{eq:shadowSchwarzschildcheck}
\end{equation}
For the finite observer screen of Eq.~\eqref{eq:shadowfinitescreen}, its exact radius instead satisfies
\begin{equation}
\rho_{\mathrm o}^2=
\frac{27\mu^2(1-2\mu/R_{\mathrm o})}
{1-27\mu^2(1-2\mu/R_{\mathrm o})/R_{\mathrm o}^2}.
\label{eq:shadowfiniteSchwarzschildcheck}
\end{equation}
A vanishing rotation parameter with $\epsilon\ne0$ leaves the spatial orientation of the vacuum tensor in the physical geometry. Its contour is determined by the same photon calculation, with the circular Schwarzschild result serving as the joint $\chi=\epsilon=0$ limit.


\subsection{Numerical illustrations}

The following figures use $R_{\mathrm o}=1000\mu$, $128$ independently calculated directions around each contour, and a screen transition tolerance of $0.002\mu$. All displayed parameters lie inside the sufficient exterior causal domain of Sec.~\ref{sec:generalhorizon}. The normalized coordinates on both screen axes have identical ranges. The calculation checks the null Hamiltonian, the conserved axial momentum, and the observer's orthonormal frame; refinement of the screen and observer distance provides a further convergence test. For the representative case $\chi=0.8$, $\epsilon=0.08$, and $\theta_{\mathrm o}=30^\circ$, doubling the angular samples and halving the transition tolerance changes the area radius by approximately $0.02\%$. Moving the observer to $2000\mu$ changes it by approximately $0.05\%$.

\begin{figure}[t]
\centering
\begin{minipage}[t]{0.49\linewidth}\centering
\includegraphics[width=\linewidth]{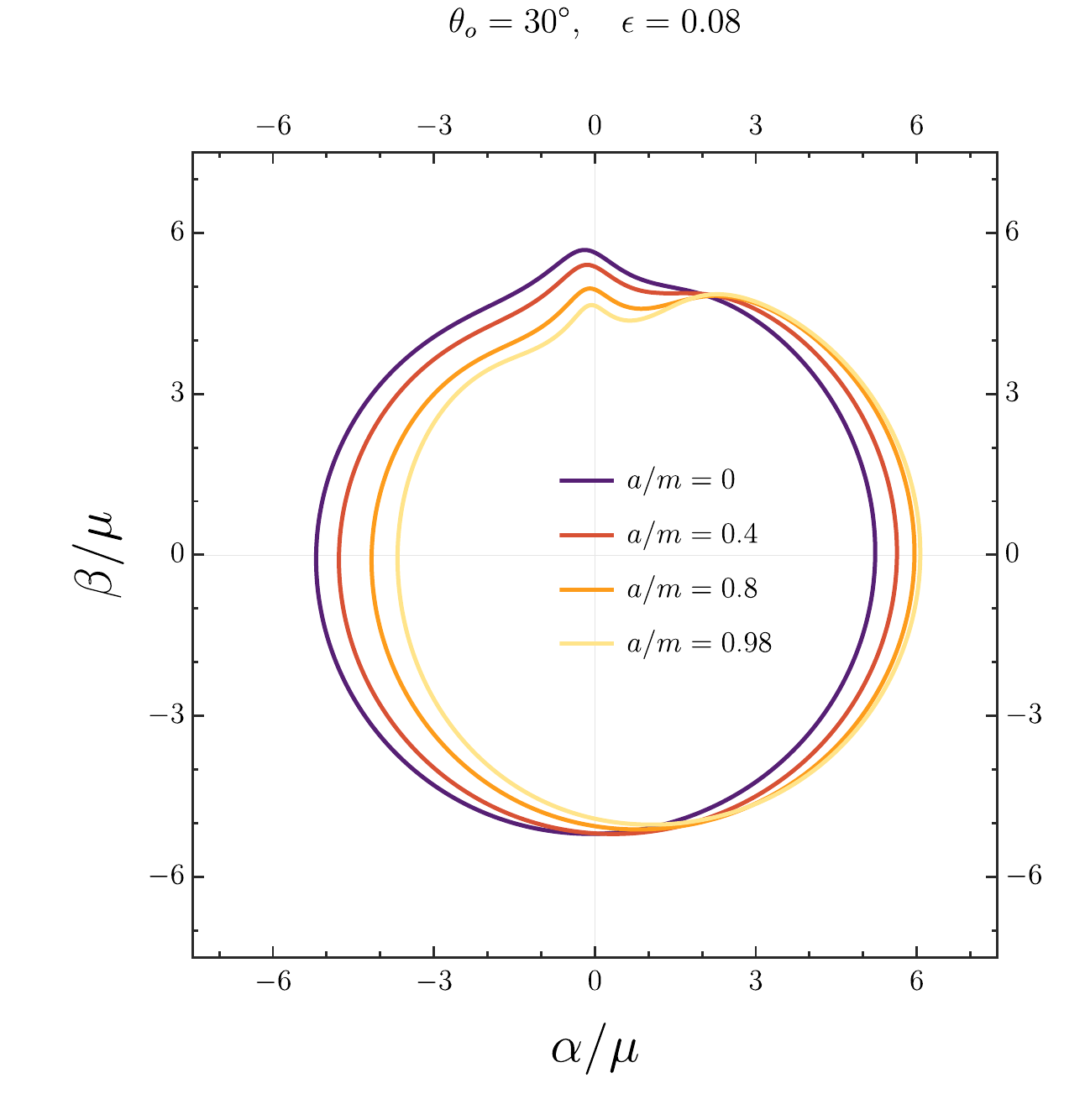}
\end{minipage}\hfill
\begin{minipage}[t]{0.49\linewidth}\centering
\includegraphics[width=\linewidth]{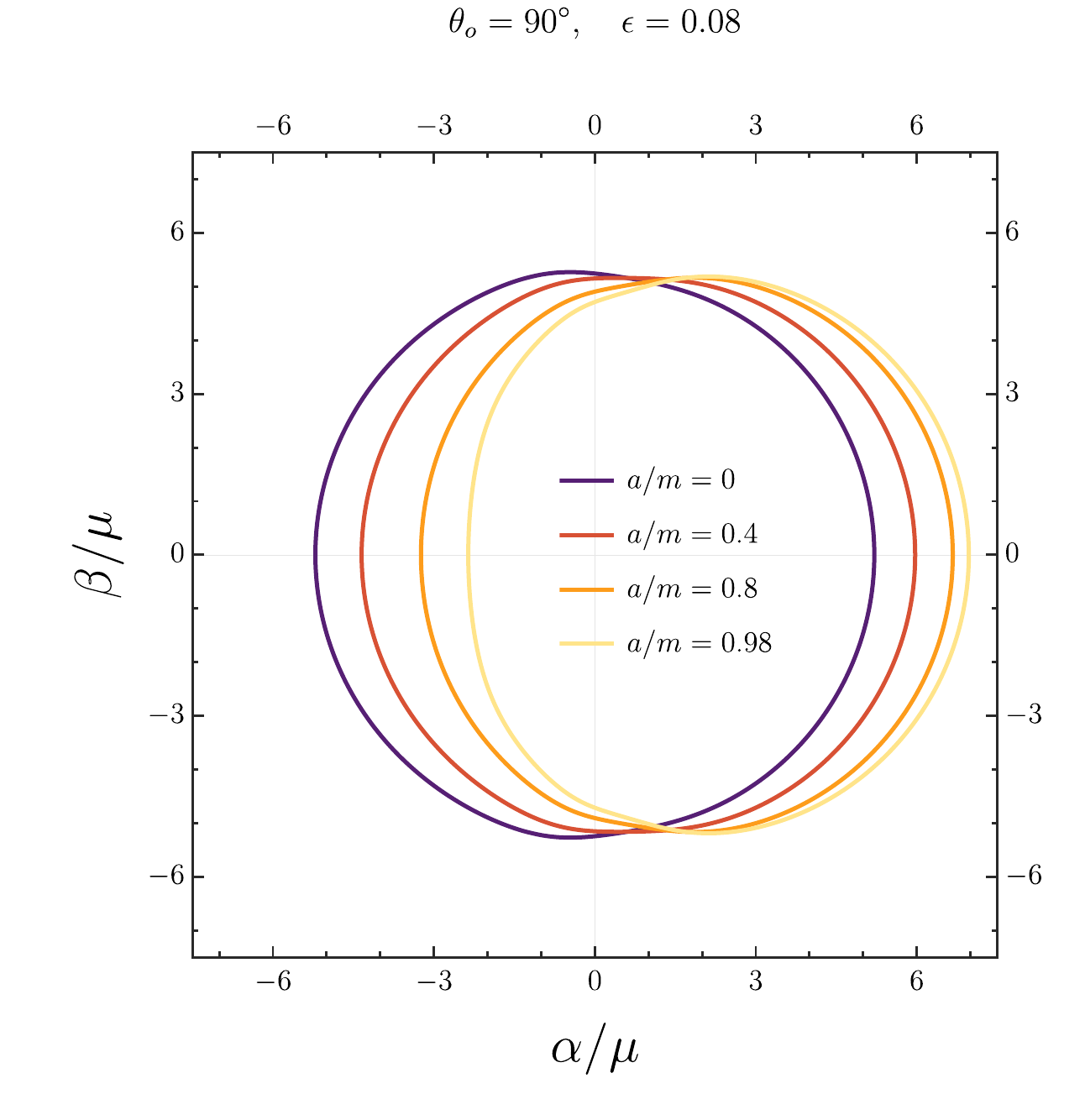}
\end{minipage}
\caption{Critical curves for $\chi=0,\,0.4,\,0.8,$ and $0.98$, with $\epsilon=0.08$. Panel (a) has $\theta_{\mathrm o}=30^\circ$ and panel (b) has $\theta_{\mathrm o}=90^\circ$. The screen coordinates are expressed in units of $\mu$. Every contour is calculated from the physical metric of Eq.~\eqref{eq:spatialphysicalmetric}.}
\label{fig:shadowrotation}
\end{figure}

Fig.~\ref{fig:shadowrotation} displays the combined effect of rotation and inclination in the chosen vacuum. Increasing rotation changes the position and shape of the contour. The equatorial view resolves the contrast between the prograde and retrograde photon trajectories, while the inclined view includes the coupled polar motion. The same normalization permits a direct comparison between the two images.

\begin{figure}[t]
\centering
\begin{minipage}[t]{0.49\linewidth}\centering
\includegraphics[width=\linewidth]{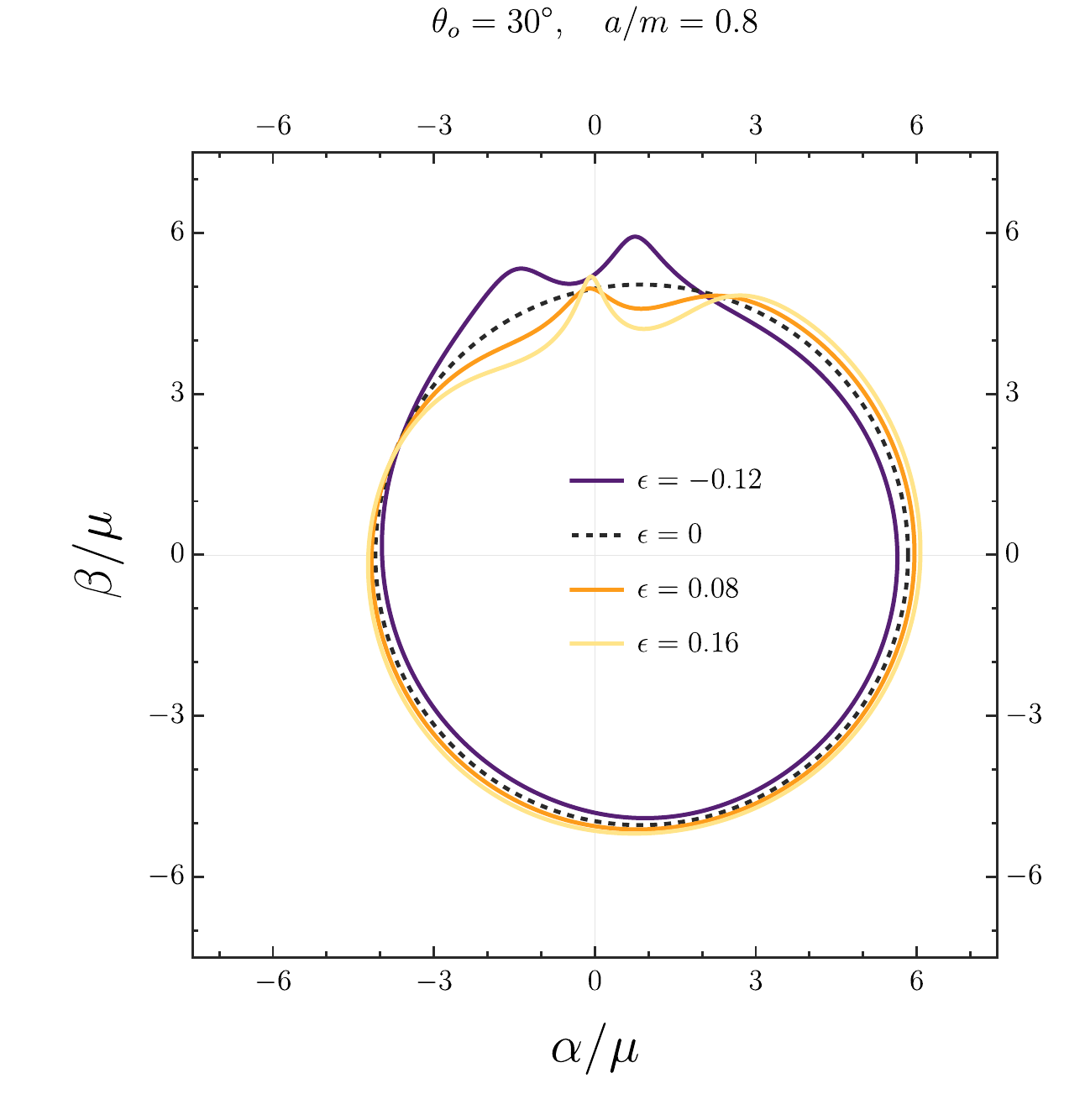}
\end{minipage}\hfill
\begin{minipage}[t]{0.49\linewidth}\centering
\includegraphics[width=\linewidth]{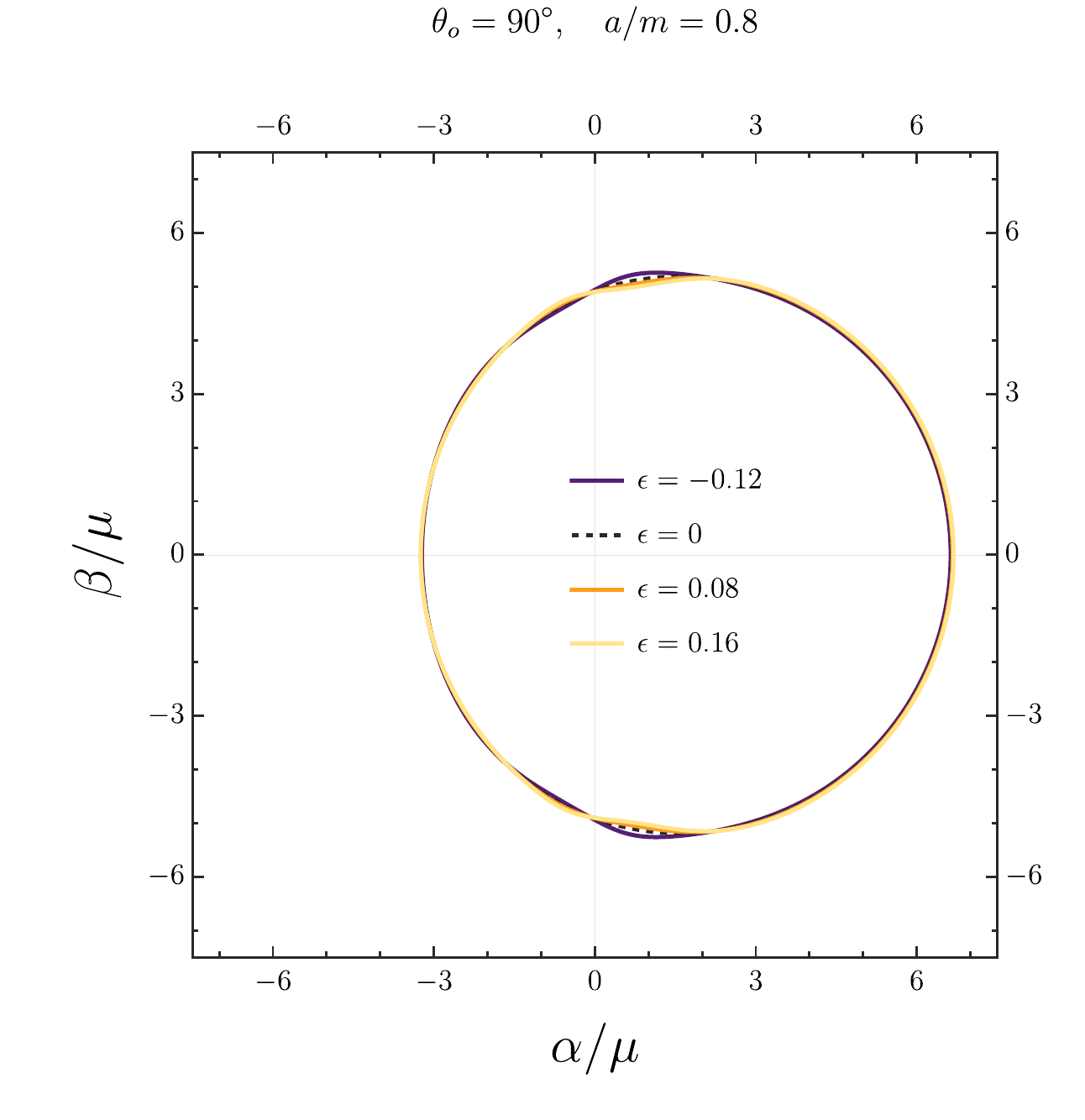}
\end{minipage}
\caption{Critical curves for $\epsilon=-0.12,\,0,\,0.08,$ and $0.16$, with $\chi=0.8$ and fixed normalized mass scale. Panel (a) has $\theta_{\mathrm o}=30^\circ$ and panel (b) has $\theta_{\mathrm o}=90^\circ$. The $\epsilon=0$ contour supplies the Kerr reference. }
\label{fig:shadowLV}
\end{figure}

Fig.~\ref{fig:shadowLV} isolates the dependence on $\epsilon$. The deformation changes the reconstructed metric and its photon dynamics at fixed $\mu$ and $\chi$. The resulting contours show us a direct geometric setting of the combination in Eq.~\eqref{eq:shadownormalizedparameters}. Their interpretation uses the complete image and the same diameter convention in every comparison.

The area radii and the centroid of the inclined image are summarized in Table~\ref{tab:shadowgeometry}. In this example, the area varies more weakly than the contour near the polar viewing directions. The inclined image resolves a vertical displacement of its centroid, while equatorial reflection gives $\beta_{\mathrm c}=0$ for the equatorial observer. These complementary measures describe the size and shape information available in the image.

\begin{table}[t]
\centering
\begin{tabular}{c c c c c}
\hline
$\epsilon$ & $\rho_{\mathrm A}(30^\circ)/\mu$ & $\rho_{\mathrm A}(90^\circ)/\mu$ & $\alpha_{\mathrm c}(30^\circ)/\mu$ & $\beta_{\mathrm c}(30^\circ)/\mu$\\
\hline
-0.12 & 4.951 & 5.069 & 0.755 & 0.208 \\
0.00 & 5.001 & 5.074 & 0.868 & 0.000 \\
0.08 & 5.031 & 5.077 & 0.936 & -0.115 \\
0.16 & 5.057 & 5.080 & 1.001 & -0.216 \\
\hline
\end{tabular}
\caption{Rounded geometric measures for the configurations of Fig.~\ref{fig:shadowLV}, with $\chi=0.8$ and $R_{\mathrm o}=1000\mu$. The centroid is defined by the enclosed image area.}
\label{tab:shadowgeometry}
\end{table}


\subsection{Comparison with Sgr A* and M87*}

An observational comparison relates the measured emission ring to a shadow diameter through an image. Schematically,
\begin{equation}
\widehat d_{\mathrm{ring}}=\mathfrak c\,d_{\mathrm{sh}},
\qquad
d_{\mathrm{sh}}=6\sqrt3\,\frac\mu{\mathcal D_{\mathrm o}}\,
f_{\mathrm{sh}}(\chi,\theta_{\mathrm o},\epsilon).
\label{eq:shadowphysicaldiameter}
\end{equation}
The factor $\mathfrak c$ includes the emitting plasma and measurement procedure. The dimensionless function $f_{\mathrm{sh}}$ is evaluated from the numerical contour with the centering and diameter estimator used in that configuration. The area diameter of Eq.~\eqref{eq:shadowgeneralarea} shows us a geometric measure; a median diameter uses its own estimator. Both give $f_{\mathrm{sh}}(0,\theta_{\mathrm o},0)=1$ in the distant Schwarzschild limit.

If $\mu_{\mathrm{dyn}}$ denotes an independently dynamical mass parameter with dimensions of length, the fractional size comparison is
\begin{equation}
\delta_{\mathrm{sh}}=
\frac{d_{\mathrm{sh}}}{6\sqrt3\,\mu_{\mathrm{dyn}}/\mathcal D_{\mathrm o}}-1,
\qquad
1+\delta_{\mathrm{sh}}=\frac\mu{\mu_{\mathrm{dyn}}}\,
f_{\mathrm{sh}}(\chi,\theta_{\mathrm o},\epsilon).
\label{eq:shadowphysicaldeviation}
\end{equation}
This relation keeps the geometric deformation and the dynamical mass explicit.

The published Sgr A* metric test gives a fiducial diameter
$d_{\mathrm{sh}}=48.9^{+5.2}_{-5.1}\,\mu\mathrm{as}$ and the fractional deviations
\begin{equation}
\delta_{\mathrm{sh}}^{\mathrm{VLTI}}=-0.08^{+0.09}_{-0.09},\qquad
\delta_{\mathrm{sh}}^{\mathrm{Keck}}=-0.04^{+0.09}_{-0.10},
\label{eq:Sgrshadowconstraints}
\end{equation}
for its two dynamical priors, at $68\%$ credibility~\cite{EHTSgr2022VI}. Their interval endpoints correspond to the 
\begin{equation}
0.83\leq\frac\mu{\mu_{\mathrm{dyn}}}f_{\mathrm{sh}}\leq1.01
\quad\text{(VLTI)},\qquad
0.86\leq\frac\mu{\mu_{\mathrm{dyn}}}f_{\mathrm{sh}}\leq1.05
\quad\text{(Keck)}.
\label{eq:Sgrshadowbenchmarks}
\end{equation}
Using these settings for the present metric needs a matched emission configuration and diameter estimator. They, in other words, define a comparison in $(\epsilon,\chi,\theta_{\mathrm o},\mu/\mu_{\mathrm{dyn}})$.

For M87*, the published comparison of the EHT gravitational angular scale with the stellar dynamics scale gives
\begin{equation}
\delta_{\mathrm g}^{\mathrm{M87}}
=\frac{\theta_{\mathrm g}^{\mathrm{EHT}}}{\theta_{\mathrm g}^{\mathrm{dyn}}}-1
=-0.01\pm0.17
\label{eq:M87shadowconstraints}
\end{equation}
at $68\%$ credibility~\cite{EHTM872019VI}. Its use for Eq.~\eqref{eq:spatialphysicalmetric} likewise calls for an image model in this geometry and for the corresponding dynamical prior. An emission ring diameter and a geometric shadow diameter have distinct configurations in this comparison.

The normalized photon equations identify the coupling combination accessible to this analysis,
\begin{equation}
\frac{\xi_2b^2}{1+\xi_1b^2}=2\epsilon.
\label{eq:shadowcouplingcombination}
\end{equation}
At fixed normalized mass and spin, the contours can probe $\epsilon$. Separate determinations of the two couplings require complementary information fixing the remaining parameter freedom. A quantitative posterior combines the photon contour, the emission model, the observer geometry, and the dynamical mass and distance. In this manner, the regular exterior solution gives us a definite geometric calculation and a route from its photon dynamics to observational constraints.


\section{Observational estimates and parameter degeneracy}
\label{sec:bounds}

The regular corotating configuration provides a definite physical geometry for the orbital and photon calculations developed in Secs.~\ref{sec:geodesics} and~\ref{sec:shadows}. Its deformation affects these predictions after the overall mass scale has been normalized. Observational information can therefore be organized around the physical parameters of this geometry, the configuration of the measured directions, and the correlations introduced by the source and observer models. We work with the smooth horizon and the admissible exterior of Sec.~\ref{sec:generalhorizon}, retaining the asymptotic qualification established there.


\subsection{Physical parameters and the coupling degeneracy}

Now, let us introduce the dimensionless couplings $\ell_1=\xi_1b^2$ and $\ell_2=\xi_2b^2$. The normalized parameters are
\begin{equation}
\mu=\frac{m}{\sqrt{1+\ell_1}},\qquad
\chi=\frac{a}{m},\qquad
\epsilon=\frac{\ell_2}{2(1+\ell_1)},\qquad
\gamma=\frac{C}{A}=1+\epsilon.
\nonumber
\end{equation}
Here, $\mu$ is the length scale of the normalized metric, and its relation to a dynamical mass is determined by the observational model. For the fixed vacuum profile used throughout this work, the physical metric in units of $\mu$ depends on $(\chi,\epsilon)$. Notice again that the Kerr geometry is recovered at $\epsilon=0$, while finite $\epsilon$ gives the orbital and image dependence described in the preceding sections.

This parametrization also identifies an exact degeneracy. For any $A_\star>0$, the family
\begin{equation}
\begin{gathered}
\ell_{1\star}=A_\star-1,\qquad
\ell_{2\star}=2\epsilon A_\star,\qquad
C_\star=A_\star(1+\epsilon),\\
m_\star=\mu\sqrt{A_\star},\qquad
a_\star=\chi\mu\sqrt{A_\star}
\end{gathered}
\label{eq:KRobservationaldegeneracy}
\end{equation}
produces the same normalized physical metric with the same azimuthal identification. Neutral orbital and electromagnetic observations can constrain $\mu$, $\chi$, and $\epsilon$, together with distance and orientation, while this family preserves their geometric predictions. An independent determination of either curvature coupling calls for information that fixes the remaining freedom, such as an independent configuration of the auxiliary parameters or a direct probe of the vacuum sector.

The exterior conditions supply geometric priors for such an analysis. In particular, $\epsilon>-1$, $|\chi|<1$, and the sufficient domain obtained in Sec.~\ref{sec:generalhorizon} can be expressed as
\begin{equation}
|\epsilon|<\left[2+
\frac{16\chi^2}{27\sqrt{1-\chi^2}}\right]^{-1}.
\label{eq:boundsadmissibleexterior}
\end{equation}
This domain ensures the exterior properties used here. Its role is to select admissible configurations before applying the data; observational likelihoods show additional information within that domain.


\subsection{Angular configuration and the physical orbital response}

The original azimuth has period $2\pi$ and gives smooth rotation axes at every finite radius. At fixed nonpolar angle, the normalized asymptotic coordinate is $\Phi=\phi/\sqrt\gamma$, with period $2\pi/\sqrt\gamma$. The limiting angular measure is 
\begin{equation}
\delta_\infty=2\pi\left[1-(1+\epsilon)^{-1/2}\right].
\label{eq:KRangulardeficit}
\end{equation}
It characterizes the fixed angle asymptotic end. The finite axes retain their regularity, and the nonuniform polar limit described in Sec.~\ref{sec:generalhorizon} remains part of the interpretation. The original $2\pi$ period is fixed by this finite radius regularity.

For the same physical equatorial orbit, define $\delta\phi$ and $\delta\Phi_{\mathrm{phys}}$ as the respective excesses over $2\pi$ in the direction of motion during a radial cycle. The latter uses the locally normalized azimuth $\Phi$. The coordinate relation gives the exact identity
\begin{equation}
\delta\phi=2\pi(\sqrt\gamma-1)
+\sqrt\gamma\,\delta\Phi_{\mathrm{phys}}.
\label{eq:exactangularadvancecorrection}
\end{equation}
The second term is calculated from the physical metric of Eq.~\eqref{eq:spatialphysicalmetric}. It contains the deformation of the orbital dynamics, including the mixed components retained in Sec.~\ref{sec:geodesics}. In this regard, angular and physical orbital response enter the comparison together. For an inclined orbit, the coupled radial and polar equations determine both the apsidal and nodal motion, and the observed directions follow from photon propagation to the observer.

An analytical illustration is given by the static equatorial limit. Set $\chi=0$ and introduce
\begin{equation}
u_R=\frac{\mu}{R},\qquad f=1-2u_R,\qquad
s_R=64u_R^6,\qquad d_R=\epsilon s_R.
\label{eq:boundsstaticdefinitions}
\end{equation}
The exact equatorial Killing block and inverse radial coefficient become
\begin{equation}
g_{TT}=-f,\qquad g_{T\Phi}=0,\qquad
g_{\Phi\Phi}=R^2(1+d_R),\qquad
g^{RR}=f(1+d_R).
\label{eq:boundsstaticequatorialmetric}
\end{equation}
The radial azimuthal component is included through the radial Schur complement. Its contribution to the cyclic angle cancels over a complete radial cycle, as in Sec.~\ref{sec:geodesics}.

For a timelike circular orbit, the azimuthal frequency and the ratio of the radial epicyclic frequency to it are
\begin{align}
\Omega_{\Phi,T}^{\,2}&=\frac{\mu}{R^3(1-2d_R)},\nonumber\\
\frac{\omega_{r,T}^{\,2}}{\Omega_{\Phi,T}^{\,2}}
&=1-6u_R-(25-54u_R)d_R+(10-12u_R)d_R^2
\equiv\mathcal K_r.
\label{eq:boundsstaticfrequencies}
\end{align}
The circular orbit is timelike when
\begin{equation}
\mathcal N_{\mathrm{circ}}
=f-R^2(1+d_R)\Omega_{\Phi,T}^{\,2}
=\frac{1-3u_R-(2-3u_R)d_R}{1-2d_R}>0,
\qquad 1-2d_R>0.
\label{eq:boundsstatictimelikecondition}
\end{equation}
Radial stability further requires $\mathcal K_r>0$. The corresponding advance in the limit of small eccentricity is
\begin{equation}
\delta\phi=2\pi\left(\frac{\sqrt{1+\epsilon}}{\sqrt{\mathcal K_r}}-1\right),
\qquad
\delta\Phi_{\mathrm{phys}}=2\pi\left(\frac1{\sqrt{\mathcal K_r}}-1\right).
\label{eq:boundsstaticadvance}
\end{equation}
These expressions display a genuine local dependence on $\epsilon$. For small $|\epsilon|$, on the stable Schwarzschild domain and away from marginal stability,
\begin{equation}
\delta\Phi_{\mathrm{phys}}-\delta\Phi_{\mathrm{Schw}}
=\frac{\pi\epsilon s_R(25-54u_R)}{(1-6u_R)^{3/2}}
+O(\epsilon^2).
\label{eq:boundslocaladvancecorrection}
\end{equation}
In this static equatorial limit, the local dynamical correction begins at order $\epsilon(\mu/R)^6$, while the leading azimuthal cycle contribution in Eq.~\eqref{eq:exactangularadvancecorrection} is $\pi\epsilon$. Rotation and inclination introduce the response selected by the complete metric; their contributions are calculated from the general orbital equations.

Mass configuration enters the same comparison. A Kepler estimator based on $\Omega_{\phi,T}$ and the coordinate $R$ would assign
\begin{equation}
\mu_{\mathrm{fit}}\equiv R^3\Omega_{\phi,T}^{\,2}
=\frac{(1+\epsilon)\mu}{1-2d_R}.
\label{eq:modifiedKeplerCalibration}
\end{equation}
An estimator based on $\Omega_{\Phi,T}$ has the corresponding locally normalized configuration. These expressions specify particular frequency estimators. A dynamical mass obtained from astrometry, spectroscopy, or ranging uses the associated angular and signal model, with the correlated parameters adjusted in the same geometry.


\subsection{Planetary precision and a conditional equatorial estimate}

Since the limiting calculation above gives a useful measure of sensitivity, let us $P_r$ be the radial period in the normalized time and $n_r=2\pi/P_r$. In the static equatorial weak field, at fixed mass scale and specified orbital parameters, the leading supplementary rate in the adopted $\phi$ convention is
\begin{equation}
\Delta\dot\varpi_\phi\simeq\frac{n_r\epsilon}{2},
\qquad
\frac{\xi_2b^2}{1+\xi_1b^2}=2\epsilon.
\label{eq:KRperihelionrate}
\end{equation}
Corrections include naturally the relativistic angular contribution, the physical metric response, and the mass. At fixed eccentricity in the weak field, their effect on the leading estimate is suppressed by powers of $\mu/p$ or $\epsilon$, where $p$ is the semilatus rectum. Eq.~\eqref{eq:KRperihelionrate} describes a specified equatorial angular response. Its application to an observed perihelion longitude requires the orientation of the vacuum axis, the orbital plane, and the reference frame to establish the relation with $\phi$, as we should expect.

The INPOP15a analysis reports supplementary Mercury intervals of $0\pm3.1$ and $0\pm1.05\,\mathrm{mas}\,\mathrm{cy}^{-1}$ for residual acceptance criteria of $35\%$ and $5\%$, respectively~\cite{Fienga2016}. These intervals inherit the acceptance prescription of that analysis. Mercury's sidereal period of $87.969$ days~\cite{NASAMercury} gives, using a Julian century of $36525$ days,
\begin{equation}
n\simeq5.38103\times10^{11}\,\mathrm{mas}\,\mathrm{cy}^{-1}.
\label{eq:Mercurymeanmotion}
\end{equation}
Using $n$ for $n_r$ at leading order and assigning a tolerance $\tau_{\varpi,\mathrm{acc}}$ to the response in Eq.~\eqref{eq:KRperihelionrate} gives the conditional scale
\begin{equation}
|\epsilon|_{\mathrm{sens}}\simeq
\frac{2\tau_{\varpi,\mathrm{acc}}}{n}.
\label{eq:generalMercurysensitivity}
\end{equation}
The numerical values are summarized in Table~\ref{tab:MercuryKRprecision}.

\begin{table}[t]
\centering
\begin{tabular}{c c c}
\hline
Residual criterion & $\tau_{\varpi,\mathrm{acc}}$ $(\mathrm{mas}\,\mathrm{cy}^{-1})$ & $|\epsilon|_{\mathrm{sens}}$\\
\hline
$35\%$ & $3.1$ & $1.15\times10^{-11}$\\
$5\%$ & $1.05$ & $3.90\times10^{-12}$\\
\hline
\end{tabular}
\caption{Conditional sensitivity of the static equatorial angular response to the historical Mercury intervals of INPOP15a. }
\label{tab:MercuryKRprecision}
\end{table}

For the larger interval, the corresponding coupling combination has the sensitivity scale
\begin{equation}
\left|\frac{\xi_2b^2}{1+\xi_1b^2}\right|_{\mathrm{sens}}
\simeq2.30\times10^{-11}.
\label{eq:conditionalKRbound}
\end{equation}
This number quantifies the limiting response under the specified assumptions. A fitted Kalb--Ramond constraint is obtained by incorporating the physical trajectories and photons into the ephemeris model and adjusting the correlated parameters against the observations~\cite{FiengaMinazzoli2024}.

The MESSENGER analysis gives Mercury's total fitted perihelion rate as $575.3100\pm0.0015$ arcseconds per century~\cite{Park2017Mercury}. Its quoted uncertainty shows an additional indication of measurement precision within that orbital model. A likelihood for $\epsilon$ uses the observations and their mass, quadrupole, orbital, and reference frame correlations. Several planetary orbits then supply complementary responses through their different orientations and radial ranges. A Solar System realization also requires a stellar interior matched to the chosen vacuum and a treatment of the interactions among the bodies; these specify how the regular black hole exterior construction is extended to that physical setting.


\subsection{Stellar motion and images of the Galactic center}

Motion around Sgr A* probes the same geometric deformation through a different set of trajectories and measured directions. The 2024 GRAVITY analysis obtains $f_{\mathrm{SP}}=0.918\pm0.128$ for S2 with the NACO reference frame priors, and $f_{\mathrm{SP}}=1.135\pm0.110$ from S2, S29, S38, and S55 together~\cite{GRAVITY2024ExtendedMass}. The Schwarzschild reference for S2 is approximately $12.1$ arcminutes per orbit. In those fits, $f_{\mathrm{SP}}$ multiplies the first post Newtonian acceleration and is determined together with the mass, distance, stellar orbital elements, and reference frame parameters.

For the present geometry, each star follows the timelike Hamiltonian of Sec.~\ref{sec:geodesics}, and the received photons determine the astrometric and spectroscopic predictions. The inclination relative to the vacuum axis belongs to this calculation. The extended mass distribution and other environmental contributions enter as additional parameters in the common fit.

The image calculation of Sec.~\ref{sec:shadows} adds sensitivity to the same $\epsilon$ at fixed normalized mass and spin. Its angular size and shape can be written schematically as
\begin{equation}
\theta_{\mathrm{sh}}=\frac{\mu}{\mathcal D_{\mathrm o}}
\,\mathcal S(\chi,\theta_{\mathrm o},\epsilon),\qquad
\mathbf h_{\mathrm{sh}}=\mathbf h(\chi,\theta_{\mathrm o},\epsilon),
\label{eq:boundsjointimageobservables}
\end{equation}
where $\mathcal S$ represents the chosen diameter estimator and $\mathbf h_{\mathrm{sh}}$ contains dimensionless shape measures. The finite observer radius is included when required. Stellar dynamics and images can refine the inferred physical scale, rotation, orientation, and deformation within one source model. The coupling family in Eq.~\eqref{eq:KRobservationaldegeneracy} continues to describe their remaining exact geometric degeneracy.


\subsection{A common fit and the interpretation of constraints}

The physical comparison is most naturally expressed through a vector of predicted measurements $\mathbf y(\epsilon,\boldsymbol\nu)$, where $\boldsymbol\nu$ collects the mass scale, spin, distance, orientation, orbital initial data, observer setting, and environmental parameters appropriate to the experiment. The prediction combines the massive trajectories, light propagation, and local measurements in the reconstructed metric~\cite{FiengaMinazzoli2024}. Near the Kerr limit,
\begin{equation}
\mathbf y(\epsilon,\boldsymbol\nu)
=\mathbf y_{\mathrm K}(\boldsymbol\nu)
+\epsilon\,\mathbf s(\boldsymbol\nu)+O(\epsilon^2),
\qquad
\mathbf s=\left.\frac{\partial\mathbf y}{\partial\epsilon}\right|_{\epsilon=0}.
\label{eq:boundsphysicalresponse}
\end{equation}
The response $\mathbf s$ is evaluated from the complete physical metric and the observation model. It gives the experiment specific replacement for a universal conversion of a fitted relativistic parameter into a curvature coupling.

For a local Gaussian likelihood with data covariance $\mathsf N$, let $\mathsf J=\partial\mathbf y_{\mathrm K}/\partial\boldsymbol\nu$ and let $\mathsf P_\nu$ be the prior precision matrix of the nuisance parameters. When the nuisance block $\mathsf J^{\mathsf T}\mathsf N^{-1}\mathsf J+\mathsf P_\nu$ is invertible, marginalizing its linear response gives
\begin{equation}
\mathcal I_\epsilon=
\mathbf s^{\mathsf T}\left[\mathsf N^{-1}
-\mathsf N^{-1}\mathsf J
(\mathsf J^{\mathsf T}\mathsf N^{-1}\mathsf J+\mathsf P_\nu)^{-1}
\mathsf J^{\mathsf T}\mathsf N^{-1}\right]\mathbf s.
\label{eq:boundsmarginalinformation}
\end{equation}
When this information is positive and the linear approximation applies, $\sigma_\epsilon\simeq\mathcal I_\epsilon^{-1/2}$. The expression makes the mass, orientation, and reference frame correlations explicit. Published residual acceptance intervals maintain their own statistical prescription, while a covariance based analysis uses the corresponding measurement likelihood.

Radio propagation and redshift measurements can be included in the same construction. For example, Cassini obtained $\gamma_{\mathrm{PPN}}-1=(2.1\pm2.3)\times10^{-5}$ within its parametrized post Newtonian signal model~\cite{Bertotti2003Cassini}. Applying that precision to the present vacuum requires the appropriate weak field expansion, source matching, and radio link prediction. This procedure assigns the measured sensitivity to the physical response of the regular metric and incorporates it with the other observations.

Finally, a fitted interval $[\epsilon_-,\epsilon_+]$ translates into the coupling region
\begin{equation}
2(1+\ell_1)\epsilon_-\leq\ell_2
\leq2(1+\ell_1)\epsilon_+,
\qquad \ell_1>-1,
\label{eq:boundscouplingregion}
\end{equation}
intersected with the chosen admissible exterior domain. Additional orbital or image information can narrow this region through its dependence on $\epsilon$; an independent determination of the two couplings requires information along the remaining degeneracy. In this manner, the regular corotating solution supplies both the geometric predictions and the parameter relations needed for a consistent observational analysis.


\section{Conclusion}
\label{sec:conclusion}

We derived the field equations of metric affine Kalb--Ramond gravity and the algebraic relation between the physical and auxiliary metrics before selecting a spacetime symmetry. In the vacuum sector with zero cosmological constant, the Kerr auxiliary metric and a closed corotating two form of fixed positive norm satisfied the connection, metric, and tensor equations. The physical reconstruction was $g_{\mu\nu}=\frac{1}{C}\left(h_{\mu\nu}+\xi_2A\,b_{\mu\alpha}h^{\alpha\beta}b_{\nu\beta}\right)$.

The temporal components placed the horizon generator in the tensor kernel. Its physical metric dual was the auxiliary metric dual divided by $C$, preserving its null character and normal direction at $r_+=m+\sqrt{m^2-a^2}$. The metrics, independent connection, and vacuum tensor were smooth across this physical Killing horizon and the rotation axes at finite radius for unequal $A$ and $C$. With $\chi=a/m$ and $\epsilon=C/A-1$, the sufficient domain $|\epsilon|<\left[2+\frac{16\chi^2}{27\sqrt{1-\chi^2}}\right]^{-1}$, $|\chi|<1$, ensured a global time function in the exterior and outward causal paths to arbitrarily large radii. The normalized asymptotic azimuth had period $2\pi/\sqrt{1+\epsilon}$ at fixed nonpolar angle, while the original $2\pi$ period gave smooth axes at finite radius.

We obtained the particle and photon equations from the physical Hamiltonian, which retained the mixed metric components and the coupled radial and polar motion. Equatorial reflection supplied an invariant plane for the effective potentials, circular orbit energies and angular momenta, radial stability, and perihelion advance. For timelike circular orbits in the static equatorial limit, setting $\mu=m/\sqrt A$, $u=\mu/R$, and $d=64\epsilon u^6$ gave the exact frequency ratio $\frac{\omega_{r,T}^{\,2}}{\Omega_{\Phi,T}^{\,2}}=1-6u-(25-54u)d+(10-12u)d^2$. For nearly circular weak field orbits at small deformation, the local perihelion correction began at order $\epsilon u^6$, and the adopted azimuthal cycle convention added $\pi\epsilon$ at leading order.

The critical curves were obtained by tracing the physical null equations from the observer's orthonormal screen. Their deformation persisted at fixed $\mu$ and $\chi$. For $\chi=0.8$, $\theta_{\mathrm o}=30^\circ$, and $R_{\mathrm o}=1000\mu$, changing $\epsilon$ from $0$ to $0.08$ increased the area radius from $5.001\mu$ to $5.031\mu$ and shifted the vertical centroid from $0$ to $-0.115\mu$. Equatorial reflection preserved a zero vertical centroid for the equatorial observer. The equal factor limit reproduced the Kerr metric and its critical curve.

The observational comparison isolated $\epsilon=\xi_2b^2/[2(1+\xi_1b^2)]$. With $\ell_i=\xi_i b^2$, the relation $\ell_2=2\epsilon(1+\ell_1)$ defined the exact coupling degeneracy at fixed $(\mu,\chi,\epsilon)$, with $m=\mu\sqrt{1+\ell_1}$ and $a=\chi m$. Under the static equatorial angular convention and mass setting used here, Mercury's $3.1\,\mathrm{mas}\,\mathrm{cy}^{-1}$ tolerance from the $35\%$ INPOP15a residual criterion gave $\left|\frac{\xi_2b^2}{1+\xi_1b^2}\right|_{\mathrm{sens}}\simeq2.30\times10^{-11}$. 


\section*{Acknowledgments}
\hspace{0.5cm} A. A. Araújo Filho is supported by Conselho Nacional de Desenvolvimento Cient\'{\i}fico e Tecnol\'{o}gico (CNPq) -- [150223/2025-0]. The author thanks the organizers of the first edition of the Simpósio Paraibano de Gravitação e Teoria de Campos for the invitation to present the results of this paper, and the audience for the fruitful discussions and suggestions.

\section*{Data Availability Statement}

Data associated with this study consist of the analytical expressions and numerical figures presented in the manuscript. No additional dataset is required to reproduce the analytical results. The code used to generate the numerical results and figures is available from the corresponding author upon reasonable request.

\bibliographystyle{apsrev4-2}
\bibliography{main}

\end{document}